\documentclass{amsart}

\RequirePackage{amsmath}
\RequirePackage{bm}
\RequirePackage{amssymb}
\RequirePackage{upref}
\RequirePackage{amsthm}
\RequirePackage{enumerate}
\RequirePackage{pb-diagram}
\RequirePackage{amsfonts}
\RequirePackage[mathscr]{eucal}
\RequirePackage{verbatim}
\RequirePackage{xr}
\RequirePackage{graphicx}
\usepackage{calc}
\usepackage{xspace}
\RequirePackage{color}
\RequirePackage{ifthen}

\newcommand{\cf}{cf.\@\xspace}
\newcommand{\resp}{resp.\@\xspace}

\newcommand{\al}{\alpha}
\newcommand{\bet}{\beta}
\newcommand{\ga}{\gamma}
\newcommand{\de}{\delta }
\newcommand{\e}{\epsilon}

\newcommand{\f}{\varphi}
\newcommand{\h}{\eta}

\newcommand{\lam}{\lambda}

\newcommand{\n}{\nu}
\newcommand{\om}{\omega}

\newcommand{\s}{\sigma}
\newcommand{\x}{\xi}

\newcommand{\C}{\varGamma}
\newcommand{\D}{\varDelta}
\newcommand{\F}{\varPhi}
\newcommand{\Lam}{\varLambda}
\newcommand{\Om}{\varOmega}

\newcommand{\di}[1]{#1\nobreakdash-\hspace{0pt}dimensional}

\newcommand{\fv}[2]{#1\hspace{0pt}_{|_{#2}}}

\newcommand{\so}{{\mc S_0}}

\newcommand{\const}{\tup{const}}
\newcommand{\ndash}{\nobreakdash--}

\newcommand{\msp[1]}[1]{\mspace{#1mu}}

\newcommand{\R}[1][n+1]{{\protect\mathbb R}^{#1}}

\newcommand{\Cc}{{\protect\mathbb C}}

\newcommand{\N}{{\protect\mathbb N}}

\newcommand{\eR}{\stackrel{\lower1ex \hbox{\rule{6.5pt}{0.5pt}}}{\msp[3]\R[]}}
\newcommand{\eN}{\stackrel{\lower1ex \hbox{\rule{6.5pt}{0.5pt}}}{\msp[1]\N}}
\newcommand{\eO}{\stackrel{\lower1ex \hbox{\rule{6pt}{0.5pt}}}{\msc O}}

\DeclareMathOperator{\graph}{graph}

\DeclareMathOperator{\supp}{supp}
\DeclareMathOperator{\id}{id}

\DeclareMathOperator{\imc}{Im}

\newcommand\ra{\rightarrow}

\newcommand\hra{\hookrightarrow}

\newcommand{\rha}{\rightharpoondown}

\newcommand\pa{\partial}
\newcommand\pde[2]{\frac {\partial#1}{\partial#2}}
 
\newcommand\dde[2]{\frac {\de#1}{\de#2}}

\newcommand{\un}{\infty}
\newcommand{\A}{\forall}

\newcommand{\set}[2]{\{\,#1\colon #2\,\}}
\newcommand{\uu}{\cup}
\newcommand{\ii}{\cap}
\newcommand{\uuu}{\bigcup}

\newcommand{\uud}{ \stackrel{\lower 1ex \hbox {.}}{\uu}}
\newcommand{\uuud}[1]{ \stackrel{\lower 1ex \hbox {.}}{\uuu_{#1}}}
\newcommand\su{\subset}

\newcommand{\sminus}[1][28]{\raise 0.#1ex\hbox{$\scriptstyle\setminus$}}

\newcommand{\wed}{\wedge}

\newcommand{\abs}[1]{\lvert#1\rvert}

\newcommand{\norm}[1]{\lVert#1\rVert}

\newcommand{\spd}[2]{\protect\langle #1,#2\protect\rangle}

\newcommand\ch[3]{\varGamma_{#1#2}^#3}
\newcommand\cha[3]{{\bar\varGamma}_{#1#2}^#3}

\newcommand{\riem}[4]{R_{#1#2#3#4}}
\newcommand{\riema}[4]{{\bar R}_{#1#2#3#4}}

\newcommand{\rnn}{\bar R_{\al\bet}\nu^\al\nu^\bet}
\newcommand{\ein}{G_{\al\bet}\nu^\al\nu^\bet}

\newcommand{\tit}{\textit}

\newcommand{\tup}{\textup}

\newcommand{\mc}{\protect\mathcal}
\newcommand{\msc}{\protect\mathscr}

\providecommand{\bysame}{\makebox[3em]{\hrulefill}\thinspace}

\newcommand{\cq}[1]{\glqq{#1}\grqq\,}

\newcommand{\bt}{\begin{thm}}
\newcommand{\bl}{\begin{lem}}
\newcommand{\bc}{\begin{cor}}
\newcommand{\bd}{\begin{definition}}
\newcommand{\bpp}{\begin{prop}}
\newcommand{\br}{\begin{rem}}
\newcommand{\bn}{\begin{note}}
\newcommand{\be}{\begin{ex}}
\newcommand{\bes}{\begin{exs}}
\newcommand{\bb}{\begin{example}}
\newcommand{\bbs}{\begin{examples}}
\newcommand{\ba}{\begin{axiom}}
\newcommand{\bas}{\begin{assumption}}

\newcommand{\et}{\end{thm}}
\newcommand{\el}{\end{lem}}
\newcommand{\ec}{\end{cor}}
\newcommand{\ed}{\end{definition}}
\newcommand{\epp}{\end{prop}}
\newcommand{\er}{\end{rem}}
\newcommand{\en}{\end{note}}
\newcommand{\ee}{\end{ex}}
\newcommand{\ees}{\end{exs}}
\newcommand{\eb}{\end{example}}
\newcommand{\ebs}{\end{examples}}
\newcommand{\ea}{\end{axiom}}
\newcommand{\eas}{\end{assumption}}

\newcommand{\bp}{\begin{proof}}
\newcommand{\ep}{\end{proof}}
\newcommand{\eps}{\renewcommand{\qed}{}\end{proof}}

\newcommand{\bal}{\begin{align}}

\newcommand{\bi}[1][1.]{\begin{enumerate}[\upshape #1]}
\newcommand{\bia}[1][(1)]{\begin{enumerate}[\upshape #1]}
\newcommand{\bin}[1][1]{\begin{enumerate}[\upshape\bfseries #1]}
\newcommand{\bir}[1][(i)]{\begin{enumerate}[\upshape #1]}
\newcommand{\bic}[1][(i)]{\begin{enumerate}[\upshape\hspace{2\cma}#1]}
\newcommand{\bis}[2][1.]{\begin{enumerate}[\upshape\hspace{#2\parindent}#1]}
\newcommand{\ei}{\end{enumerate}}

\newcommand\ndots{\raise 0.47ex \hbox {,}\hskip0.06em\cdots %
     \raise 0.47ex \hbox {,}\hskip0.06em} 

\newcommand{\q}{\quad}
\newcommand{\qq}{\qquad}

\newcommand{\hp}{\hphantom}

\newcommand\nd{\noindent}

\newskip\Csmallskipamount                                                
\Csmallskipamount=\smallskipamount
\newskip\Cmedskipamount
\Cmedskipamount=\medskipamount
\newskip\Cbigskipamount
\Cbigskipamount=\bigskipamount

\newcommand\cvs{\vspace\Csmallskipamount}   
\newcommand\cvm{\vspace\Cmedskipamount}

\newskip\csa
\csa=\smallskipamount

\newskip\cma
\cma=\medskipamount

\newskip\cba
\cba=\bigskipamount

\newdimen\spt
\newcommand\citem{\cvs\advance\itemno by
1{(\romannumeral\the\itemno})\hskip3pt}
\newcommand{\bitem}{\cvm\nd\advance\itemno by
1{\bf\the\itemno}\hspace{\cma}}

\newcount\itemno
\newcommand{\las}[1]{\label{S:#1}}

\newcommand{\lae}[1]{\label{E:#1}}
\newcommand{\lat}[1]{\label{T:#1}}
\newcommand{\lal}[1]{\label{L:#1}}

\newcommand{\lac}[1]{\label{C:#1}}

\newcommand{\lar}[1]{\label{R:#1}}

\newcommand{\rs}[1]{Section~\ref{S:#1}}

\newcommand{\rt}[1]{Theorem~\ref{T:#1}}
\newcommand{\rl}[1]{Lemma~\ref{L:#1}}

\newcommand{\re}[1]{\eqref{E:#1}}
\newcommand{\frc}[1]{Corollary~\ref{C:#1} on page~\tup{\pageref{C:#1}}}
\newcommand{\frt}[1]{Theorem~\ref{T:#1} on page~\tup{\pageref{T:#1}}}
\newcommand{\frl}[1]{Lemma~\ref{L:#1} on page~\tup{\pageref{L:#1}}}
\newcommand{\frr}[1]{Remark~\ref{R:#1} on page~\tup{\pageref{R:#1}}}

\newcommand{\fre}[1]{\eqref{E:#1} on page~\tup{\pageref{E:#1}}}

\newskip\thmskip
\thmskip=\parindent

\newskip\hsk
\newenvironment{hinw}{\labelsep=0pt\begin{list}{}{\labelsep=0pt\itemindent=0pt\labelwidth=0pt\leftmargin=\parindent\rightmargin=0pt\partopsep=\cba}%
\item\it\nopagebreak\nopagebreak}%
{\end{list}}

\newcommand\bh{\begin{hinw}}
\newcommand{\eh}{\end{hinw}}

\newtheoremstyle{normal}
  {\cba}
  {\cba}
  {}
  {\thmskip}
  {\bfseries}
  {.}
  {\hsk}
  {}

\newtheoremstyle{abschnitt}
  {\cba}
  {\cba}
  {}
  {\thmskip}
  {\bfseries}
  {.}
  {\hsk}
  {}

\newtheoremstyle{italic}
  {\cba}
  {\cba}
  {\itshape}
  {\thmskip}
  {\bfseries}
  {.}
  {\hsk}
  {}

\newtheoremstyle{aufgaben}
  {\cba}
  {\cba}
  {}
  {}
  {\normalsize\bfseries}
  {.}
  {\hsk}
  {}

\newtheoremstyle{break}
  {\cba}
  {\cba}
  {\itshape}
  {}
  {\bfseries}
  {.}
  {\newline}
  {}

\theoremstyle{italic}
\newtheorem{thm}[subsection]{Theorem}
\newtheorem{lem}[subsection]{Lemma}
\newtheorem{prop}[subsection]{Proposition}
\newtheorem{cor}[subsection]{Corollary}

\theoremstyle{normal}
\newtheorem{rem}[subsection]{Remark}
\newtheorem{definition}[subsection]{Definition}
\newtheorem{example}[subsection]{Example}
\newtheorem{examples}[subsection]{Examples}
\newtheorem{ex}[subsection]{Exercise}
\newtheorem{note}[subsection]{}
\newtheorem{axiom}[subsection]{Axiom}
\newtheorem{assumption}[subsection]{Assumption}

\theoremstyle{aufgaben}
\newtheorem{exs}[subsection]{Exercises}

\numberwithin{equation}{section}
\numberwithin{figure}{section}

\newenvironment{textequation}[1][0.8]
{\begin{equation}
\begin{aligned}
\begin{minipage}{#1\linewidth}}
{\end{minipage}
\end{aligned}
\end{equation}
\ignorespacesafterend}

\newcommand{\btext}{\begin{textequation}}
\newcommand{\etext}{\end{textequation}}

\def\hinweis{\@startsection{subsection}{2}%
 \z@{0.7\linespacing\@plus 0.5\linespacing}{0.7\linespacing}%
{\normalfont\itshape\indent}}

\newcounter{hours}\newcounter{minutes}
\newcommand{\printtime}{%
\setcounter{hours}{\time/60}%
\setcounter{minutes}{\time-\value{hours}*60}%
\ifthenelse{\value{minutes}<10}{\thehours :0\theminutes}{\thehours:\theminutes}}

\usepackage[german,english]{babel}
\usepackage{graphicx}
\RequirePackage{amsmath}
\RequirePackage{bm}
\RequirePackage{amssymb}
\RequirePackage{upref}
\RequirePackage{amsthm}
\RequirePackage{enumerate}
\RequirePackage{pb-diagram}
\RequirePackage{amsfonts}
\RequirePackage[mathscr]{eucal}
\RequirePackage{verbatim}
\RequirePackage{xr}
\RequirePackage{graphicx}
\usepackage{calc}
\usepackage{xspace}
\usepackage{dsfont}

\makeatletter
\RequirePackage{color}
\newcommand{\ann}[1]{\renewcommand{\@makefnmark}{\mbox{$^{\color{red}{\@thefnmark}}$}}%
\footnote {#1}}
\makeatother

\RequirePackage{upref}
\RequirePackage{amsthm}
\RequirePackage{enumerate}
\usepackage[mathscr]{eucal}

\usepackage{xr-hyper}

\usepackage{calc}

\newlength{\oddsidemarginlength}
\newlength{\topmarginlength}

\newcounter{numberoflines}
\newcounter{tempcc}
\oddsidemargin=\oddsidemarginlength
\evensidemargin=\oddsidemargin
\topmargin=\topmarginlength
\usepackage[colorlinks=true,linkcolor=blue,citecolor=blue,urlcolor=blue]{hyperref}  

\begin{document}

\flushbottom


\title[The quantization of gravity]{A unified field theory I: The quantization of gravity}

\author{Claus Gerhardt}
\address{Ruprecht-Karls-Universit\"at, Institut f\"ur Angewandte Mathematik,
Im Neuenheimer Feld 205, 69120 Heidelberg, Germany}
\email{\href{mailto:gerhardt@math.uni-heidelberg.de}{gerhardt@math.uni-heidelberg.de}}
\urladdr{\href{http://www.math.uni-heidelberg.de/studinfo/gerhardt/}{http://www.math.uni-heidelberg.de/studinfo/gerhardt/}}

%
\date{\today}
%


\begin{abstract} 
In a former paper we proposed a model for the quantization of gravity by working in a bundle $E$ where we  realized the Hamilton constraint as the Wheeler-DeWitt equation. However, the corresponding operator only acts in the fibers and not in the base space. Therefore, we now discard the Wheeler-DeWitt equation and express the Hamilton constraint differently, either with the help of the Hamilton equations or by employing a geometric evolution equation. There are two  modifications possible which both are equivalent to the Hamilton constraint and which lead to two new models. In the first model we obtain a hyperbolic operator that acts in the fibers as well as in the base space and we can construct a symplectic vector space and a Weyl system.

\nd
In the second model  the resulting equation is a wave equation in $\so \times (0,\infty)$ valid in points $(x,t,\xi)$ in $E$ and we look for solutions for each fixed $\xi$. This set of equations contains as a special case the  equation of a  quantized cosmological Friedmann universe without matter but with a cosmological constant, when we look for solutions which only depend on $t$. Moreover, in case $\so$ is compact we prove a spectral resolution of the equation.
\end{abstract}

\maketitle

\tableofcontents

\setcounter{section}{0}
\section{Introduction}

The quantization of gravity is hampered by the fact that the Einstein-Hilbert Lagrangian is singular. Switching to a Hamiltonian setting requires to impose two constraints, the Hamilton constraint and the diffeomorphism constraint. Though we were able to eliminate the diffeomorphism constraint in a recent paper \cite{cg:qgravity}, the Hamilton constraint is a serious obstacle. Quantization of a Hamiltonian setting requires a model in which the quantized variables, which turn into operators, act, and, in case of constraints, preferably given as an equation, to quantize this equation.

In the former paper we proposed a quantization of gravity by working in a fiber bundle $E$ with base space $\so$ after quantization, the Hamilton function $H$ was transformed to an hyperbolic operator $\hat H$ and the Hamilton condition, which could be expressed by
\begin{equation}\lae{1.1}
H=0,
\end{equation}
was transformed to the Wheeler-DeWitt equation
\begin{equation}
\hat H u=0
\end{equation}
in the bundle $E$. However, the operator $\hat H$ acts only in the fibers, there is no differentiation in the base space $\so$, though the solutions are defined in $E$. This seems to be unsatisfactory.

In this paper we want to offer a better quantization model: We are still working in the bundle $E$, but we discard the Wheeler-DeWitt equation, i.e., we do not express the Hamilton constraint by equation \re{1.1} but differently using the Hamilton equations. The second Hamilton equation has the form
\begin{equation}
\dot \pi^{ij}=-\dde{\mc H}{g_{ij}},
\end{equation}
or equivalently,
\begin{equation}
\dot\pi^{ij}=\{\pi^{ij},\mc H\},
\end{equation}
where we use a Hamiltonian density at the moment. Hence we have the identity
\begin{equation}\lae{1.5}
g_{ij}\{\pi^{ij},\mc H\}=-g_{ij}\dde{\mc H}{g_{ij}}
\end{equation}
which is a scalar equation. 

The Hamilton constraint can be expressed in the form
\begin{equation}
\abs A^2-H^2=(R-2\Lam).
\end{equation}
Looking at the right-hand side of \re{1.5} the term $\abs A^2-H^2$, which will be transformed to be the main part of the hyperbolic operator, occurs on the right-hand side in two places. Replacing $\abs A^2-H^2$ on the right side by $(R-2\Lam)$ will give an equation that defines the Hamilton constraint.

We developed two models: In the first model we replaced $\abs A^2-H^2$ partially in \re{1.5}. The quantization of the modified equation  then leads to a hyperbolic equation
\begin{equation}
Pu=0
\end{equation}
in $E$, where $P$ acts in the fibers as well as in $\so$. $P$ is a symmetric operator and with the help of its Green's operator one can define a symplectic vector space and then a Weyl system, or a quantum field.

In the second model we use a geometric evolution equation to express the Hamilton constraint  by replacing $\abs A^2-H^2$ completely in the evolution equation. After quantization we then obtain a wave equation in $E$
\begin{equation}\lae{1.8}
\frac1{32}\frac{n^2}{n-1}\Ddot u
-(n-1) t^{2-\frac4n}\D u-\frac {n}2t^2(t^{-\frac4n}R-2\Lam)u=0
\end{equation}
in points $(x,t,\xi)\in E$, where a metric $g_{ij}$ in the fiber over $x\in\so$ has the form
\begin{equation}
g_{ij}=t^\frac4n\s_{ij}(x,\xi)
\end{equation}
and the Laplacian in \re{1.8} is defined with respect to $\s_{ij}$. Hence, for any $\xi$ we have a wave equation in
\begin{equation}
\so\times \R[*]_+
\end{equation}
with solutions $u=u(x,t,\xi)$. We prove that solutions of the corresponding Cauchy problems exist and are smooth in all variables.

This second model seems to be the right model since it contains the quantization of a cosmological Friedmann universe, without matter but with a cosmological constant, as a special case by choosing $\s_{ij}$ to be the metric of a space of constant curvature and by assuming $u=u(t)$. Equation \re{1.8} is in this case identical to the quantized Friedmann equation up to the last constant.

Moreover, assuming $\so$ to be compact we also prove a spectral resolution of equation \re{1.8}, by constructing a countable basis of solutions of the form
\begin{equation}
u=w(t)v(x),
\end{equation}
where $v$ is an eigenfunction of the problem
\begin{equation}\lae{1.12}
-(n-1)\D v-\frac{n}2R v=\mu v
\end{equation}
in $\so$ with $\mu>0$  and $w$ an eigenfunction of an ODE. These solutions have finite energy, \cf\fre{6.49}.

The results for the first model are proved and described in detail in \rs{4} and \rs {5}. The results for the second model are proved in \rs{6}. Here is a more formal summary of the results of the second model:
\bt
Let $(\so,\s_{ij})$ be a given connected, smooth and complete  $n$-dimensional Riemannian mani\-fold   and let 
\begin{equation}
Q=\so\times \R[*]_+
\end{equation}
be the corresponding globally hyperbolic spacetime equipped with the Lorentzian metric \re{6.17} or, if necessary, with \re{6.18}, then the hyperbolic equation
\begin{equation}
\frac1{32}\frac{n^2}{n-1}\Ddot u
-(n-1) t^{2-\frac4n}\D u-\frac {n}2 t^{2-\frac4n}Ru+nt^2\Lam u=0,
\end{equation}
where the Laplacian and the scalar curvature correspond to the metric $\s_{ij}$, describes a model for quantum gravity. If  $\so$ is compact a spectral resolution of this equation has been proved in the theorem below.
\et

\bt
Assume $n\ge 2$ and  $\so$ to be compact and let $(v,\mu)$ be a solution of the eigenvalue problem \re{1.12} with $\mu>0$, then there exist countably many solutions $(w_i,\Lam_i)$ of the implicit eigenvalue problem \re{6.33} such that
\begin{equation}
\Lam_i<\Lam_{i+1}<\cdots <0,
\end{equation}
\begin{equation}
\lim_i\Lam_i=0,
\end{equation}
and such that the functions
\begin{equation}
u_i=w_i v
\end{equation}
are solutions of the wave equations \re{1.8}. The transformed eigenfunctions
\begin{equation}
\tilde w_i(t)=w_i(\lam_i^{\frac n{4(n-1)}}t), 
\end{equation}
where
\begin{equation}
\lam_i=(-\Lam_i)^{-\frac{n-1}n}
\end{equation}
form a basis of the corresponding Hilbert space $H$ and also of $L^2(\R[*]_+,\Cc)$.
\et

\section{Definitions and notations}
The main objective of this section is to state the equations of Gau{\ss}, Codazzi,
and Weingarten for spacelike hypersurfaces $M$ in a \di {(n+1)} Lorentzian
manifold
$N$.  Geometric quantities in $N$ will be denoted by
$(\bar g_{ \al \bet}),(\riema  \al \bet \ga \de)$, etc., and those in $M$ by $(g_{ij}), 
(\riem ijkl)$, etc.. Greek indices range from $0$ to $n$ and Latin from $1$ to $n$;
the summation convention is always used. Generic coordinate systems in $N$ resp.
$M$ will be denoted by $(x^ \al)$ \resp $(\x^i)$. Covariant differentiation will
simply be indicated by indices, only in case of possible ambiguity they will be
preceded by a semicolon, i.e., for a function $u$ in $N$, $(u_ \al)$ will be the
gradient and
$(u_{ \al \bet})$ the Hessian, but e.g., the covariant derivative of the curvature
tensor will be abbreviated by $\riema  \al \bet \ga{ \de;\e}$. We also point out that
\begin{equation}
\riema  \al \bet \ga{ \de;i}=\riema  \al \bet \ga{ \de;\e}x_i^\e
\end{equation}
with obvious generalizations to other quantities.

Let $M$ be a \tit{spacelike} hypersurface, i.e., the induced metric is Riemannian,
with a differentiable normal $\n$ which is timelike.

In local coordinates, $(x^ \al)$ and $(\x^i)$, the geometric quantities of the
spacelike hypersurface $M$ are connected through the following equations
\begin{equation}\lae{01.2}
x_{ij}^ \al= h_{ij}\n^ \al
\end{equation}
the so-called \tit{Gau{\ss} formula}. Here, and also in the sequel, a covariant
derivative is always a \tit{full} tensor, i.e.

\begin{equation}
x_{ij}^ \al=x_{,ij}^ \al-\ch ijk x_k^ \al+ \cha  \bet \ga \al x_i^ \bet x_j^ \ga.
\end{equation}
The comma indicates ordinary partial derivatives.

In this implicit definition the \tit{second fundamental form} $(h_{ij})$ is taken
with respect to $\n$.

The second equation is the \tit{Weingarten equation}
\begin{equation}
\n_i^ \al=h_i^k x_k^ \al,
\end{equation}
where we remember that $\n_i^ \al$ is a full tensor.

Finally, we have the \tit{Codazzi equation}
\begin{equation}
h_{ij;k}-h_{ik;j}=\riema \al \bet \ga \de\n^ \al x_i^ \bet x_j^ \ga x_k^ \de
\end{equation}
and the \tit{Gau{\ss} equation}
\begin{equation}
\riem ijkl=- \{h_{ik}h_{jl}-h_{il}h_{jk}\} + \riema  \al \bet\ga \de x_i^ \al x_j^ \bet
x_k^ \ga x_l^ \de.
\end{equation}

Now, let us assume that $N$ is a globally hyperbolic Lorentzian manifold with a
 Cauchy surface. 
$N$ is then a topological product $I\times \mc S_0$, where $I$ is an open interval,
$\mc S_0$ is a  Riemannian manifold, and there exists a Gaussian coordinate
system
$(x^ \al)$, such that the metric in $N$ has the form 
\begin{equation}\lae{01.7}
d\bar s_N^2=e^{2\psi}\{-{dx^0}^2+\s_{ij}(x^0,x)dx^idx^j\},
\end{equation}
where $\s_{ij}$ is a Riemannian metric, $\psi$ a function on $N$, and $x$ an
abbreviation for the spacelike components $(x^i)$. 
We also assume that
the coordinate system is \tit{future oriented}, i.e., the time coordinate $x^0$
increases on future directed curves. Hence, the \tit{contravariant} timelike
vector $(\x^ \al)=(1,0,\dotsc,0)$ is future directed as is its \tit{covariant} version
$(\x_ \al)=e^{2\psi}(-1,0,\dotsc,0)$.

Let $M=\graph \fv u\so$ be a spacelike hypersurface
\begin{equation}
M=\set{(x^0,x)}{x^0=u(x),\,x\in\mc S_0},
\end{equation}
then the induced metric has the form
\begin{equation}
g_{ij}=e^{2\psi}\{-u_iu_j+\s_{ij}\}
\end{equation}
where $\s_{ij}$ is evaluated at $(u,x)$, and its inverse $(g^{ij})=(g_{ij})^{-1}$ can
be expressed as
\begin{equation}\lae{01.10}
g^{ij}=e^{-2\psi}\{\s^{ij}+\frac{u^i}{v}\frac{u^j}{v}\},
\end{equation}
where $(\s^{ij})=(\s_{ij})^{-1}$ and
\begin{equation}\lae{01.11}
\begin{aligned}
u^i&=\s^{ij}u_j\\
v^2&=1-\s^{ij}u_iu_j\equiv 1-\abs{Du}^2.
\end{aligned}
\end{equation}
Hence, $\graph u$ is spacelike if and only if $\abs{Du}<1$.

The covariant form of a normal vector of a graph looks like
\begin{equation}
(\n_ \al)=\pm v^{-1}e^{\psi}(1, -u_i).
\end{equation}
and the contravariant version is
\begin{equation}
(\n^ \al)=\mp v^{-1}e^{-\psi}(1, u^i).
\end{equation}
Thus, we have
\br Let $M$ be spacelike graph in a future oriented coordinate system. Then the
contravariant future directed normal vector has the form
\begin{equation}
(\n^ \al)=v^{-1}e^{-\psi}(1, u^i)
\end{equation}
and the past directed
\begin{equation}\lae{01.15}
(\n^ \al)=-v^{-1}e^{-\psi}(1, u^i).
\end{equation}
\er

In the Gau{\ss} formula \re{01.2} we are free to choose the future or past directed
normal, but we stipulate that we always use the past directed normal.
Look at the component $ \al=0$ in \re{01.2} and obtain in view of \re{01.15}

\begin{equation}\lae{01.16}
e^{-\psi}v^{-1}h_{ij}=-u_{ij}- \cha 000\mspace{1mu}u_iu_j- \cha 0j0
\mspace{1mu}u_i- \cha 0i0\mspace{1mu}u_j- \cha ij0.
\end{equation}
Here, the covariant derivatives are taken with respect to the induced metric of
$M$, and
\begin{equation}
-\cha ij0=e^{-\psi}\bar h_{ij},
\end{equation}
where $(\bar h_{ij})$ is the second fundamental form of the hypersurfaces
$\{x^0=\const\}$.

An easy calculation shows
\begin{equation}
\bar h_{ij}e^{-\psi}=-\tfrac{1}{2}\dot\s_{ij} -\dot\psi\s_{ij},
\end{equation}
where the dot indicates differentiation with respect to $x^0$.

\section{Combining the Hamilton equations with the Hamilton constraint}\las{3}

Let $N=N^{n+1}$ be a globally hyperbolic spacetime with metric $\bar g_{\al\bet}$. We consider the Einstein-Hilbert functional
\begin{equation}\lae{3.1}
J=\int_N(\bar R-2\Lam)
\end{equation}
with cosmological constant $\Lam$ and want to write it in a form such that the Lagrangian density is regular with respect to the variables $g_{ij}$ so that we can switch to an equivalent Hamiltonian setting for these components. Let $x^0$ be time function that will split the metric such that the metric can be expressed in the form
\begin{equation}\lae{3.2}
d\bar s^2=-w^2 (dx^0)^2+g_{ij}dx^idx^j,
\end{equation}
where $(x^i)$ are local coordinates of a coordinate slice
\begin{equation}
\so=\{x^0=\const\}
\end{equation}
and 
\begin{equation}
0<w\in C^\un(N).
\end{equation}
Let us define the level sets
\begin{equation}
M(t)=\{x^0=t\}
\end{equation}
and, assuming  $0\in x^0(N)$, set
\begin{equation}
\so=M(0).
\end{equation}
The coordinate system should also be future oriented such that $\{x^0>0\}$ is the future development of $\so$.

Let $h_{ij}$ be the second fundamental form of the slices $M(t)$ with respect to the past directed normal, i.e., the Gaussian formula looks like
\begin{equation}
x^\al_{ij}=h_{ij}\nu,
\end{equation}
where $\nu$ is the past directed normal. Then
\begin{equation}\lae{3.8}
h_{ij}=-\tfrac12 \dot g_{ij}w^{-1}
\end{equation}
and the functional \re{3.1} can be expressed in the form
\begin{equation}\lae{3.9}
J=\int_a^b\int_\Om\{\abs A^2-H^2+(R-2\Lam)\}w\sqrt g,
\end{equation}
where $\Om\su N$ is some open subset of $\R[n]$, $R$ the scalar curvature of $M(t)$,
\begin{equation}
H=g^{ij}h_{ij}
\end{equation}
 the mean curvature and
\begin{equation}
\abs A^2=h_{ij}h^{ij},
\end{equation}
\cf \cite[equ. (3.37)]{cg:qgravity}. This way of expressing the Einstein-Hilbert functional is known as the ADM approach, see \cite{adm:old}.

Let $F=F(h_{ij})$ be the scalar curvature operator
\begin{equation}
F=\tfrac12(H^2-\abs A^2)
\end{equation}
and let
\begin{equation}
F^{ij,kl}=g^{ij}g^{kl}-\tfrac12\{g^{ik}g^{jl}+g^{il}g^{jk}\}
\end{equation}
be its Hessian, then
\begin{equation}\lae{3.14}
F^{ij,kl}h_{ij}h_{kl}=2F=H^2-\abs A^2
\end{equation}
and
\begin{equation}\lae{3.15}
F^{ij}=F^{ij,kl}h_{kl}=Hg^{ij}-h^{ij}.
\end{equation}
In physics
\begin{equation}\lae{3.16}
G^{ij,kl}=-F^{ij,kl}
\end{equation}
is known as the DeWitt metric.

Combining \re{3.8} and \re{3.14} $J$ can be expressed in the form
\begin{equation}\lae{3.17}
J=\int_a^b\int_\Om\{\tfrac14 G^{ij,kl}\dot g_{ij}\dot g_{kl}w^{-2}+(R-2\Lam)\}w\sqrt g.
\end{equation}
The Lagrangian density $\mc L$ is a regular Lagrangian with respect to the variables $g_{ij}$. Define the conjugate momenta
\begin{equation}\lae{3.18}
\begin{aligned}
\pi^{ij}=\frac{\pa\mc L}{\pa g_{ij}}&=\tfrac12 G^{ij,kl}\dot g_{kl}w^{-1}\sqrt g\\
&= -G^{ij,kl}h_{kl}\sqrt g
\end{aligned}
\end{equation}
and the Hamiltonian density
\begin{equation}
\begin{aligned}
\mc H&=\pi^{ij}\dot g_{ij}-\mc L\\
&=\frac1{\sqrt g}wG_{ij,kl}\pi^{ij}\pi^{kl}-(R-2\Lam)w\sqrt g,
\end{aligned}
\end{equation}
where
\begin{equation}
G_{ij,kl}=\tfrac12\{g_{ik}g_{jk}+g_{il}g_{jk}\}-\tfrac1{n-1}g_{ij}g_{kl}
\end{equation}
is the inverse of $G^{ij,kl}$.

Let us now consider an arbitrary variation of $g_{ij}$ with compact support
\begin{equation}
g_{ij}(\e)=g_{ij}+\e\om_{ij},
\end{equation}
where $\om_{ij}=\om_{ij}(t,x)$ is an arbitrary smooth, symmetric tensor with compact support in $\Om$. The vanishing of the first variation leads to the Euler-Lagrange equations
\begin{equation}\lae{3.22}
G_{ij}+\Lam g_{ij}=0,
\end{equation}
i.e., to the tangential Einstein equations. We obtain these equations by either varying \re{3.1} or \re{3.9}.

To obtain the full Einstein equations we impose the Hamilton constraint, namely, that the Hamiltonian density vanishes, or equivalently, that the normal component of the Einstein equations is satisfied
\begin{equation}\lae{3.23}
G_{\al\bet}\nu^\al\nu^\bet-\Lam=0.
\end{equation}
We then conclude that any metric $(\bar g_{\al\bet})$ satisfying \re{3.2}, \re{3.22} as well as \re{3.23} has the property that it is a stationary point for the functional \re{3.1} in the class of metrics which can be split according to \re{3.2}. Applying then a former result \cite[Theorem 3.2]{cg:qgravity} we deduce that $\bar g_{\al\bet}$ satisfies the full Einstein equations.

The Lagrangian density $\mc L$ in \re{3.17} is regular with respect to the variables $g_{ij}$, hence the tangential Einstein equations are equivalent to the Hamilton equations
\begin{equation}\lae{3.24}
\dot g_{ij}=\frac{\de \mc H}{\de \pi^{ij}} 
\end{equation}
and
\begin{equation}\lae{3.40}
\dot\pi^{ij}=-\dde{\mc H}{g_{ij}},
\end{equation}
where the differentials on the right-hand side of these equations are variational or functional derivatives, i.e., they are the Euler-Lagrange operators of the corresponding functionals with respect to the indicated variables, in this case, the functional is
\begin{equation}
\int_\Om\mc H,
\end{equation}
where $\so$ is locally parameterized over $\Om\su\R[n]$. Occasionally we shall also write
\begin{equation}
\int_\so\mc H
\end{equation}
by considering $\so$ simply to be a parameter domain without any intrinsic volume element.

We have therefore proved:
\bt\lat{3.1}
Let $N=N^{n+1}$ be a globally hyperbolic spacetime and let the metric $\bar g_{\al\bet}$ be expressed as in \re{3.2}. Then, the metric satisfies the full Einstein equations if and only if the metric is a solution of the  Hamilton equations \re{3.24} and \re{3.40} and of the equation \re{3.23} which is equivalent to 
\begin{equation}
\mc H=0
\end{equation}
and is called the Hamiltonian constraint. These equations are equations for the variables $g_{ij}$. The function $w$ is merely part of the equations and not looked at as a variable though it is of course specified in the component $\bar g_{00}$.
\et

We define the Poisson brackets 
\begin{equation}
\{u,v\}=\dde u{g_{kl}}\dde v{\pi^{kl}}-\dde u{\pi^{kl}}\dde v{g_{kl}}
\end{equation}
and obtain
\begin{equation}\lae{4.2.1}
\{g_{ij},\pi^{kl}\}=\de^{kl}_{ij},
\end{equation}
where
\begin{equation}
\de^{kl}_{ij}=\tfrac12\{\de^k_i\de^l_j+\de^l_i\de^k_j\}.
\end{equation}
Then, the second Hamilton equation can also be expressed as
\begin{equation}\lae{3.32}
\dot \pi^{ij}=\{\pi^{ij},\mc H\}.
\end{equation}
In the next section we want to quantize this Hamiltonian setting and especially the Hamiltonian constraint. In order to achieve this we shall express the equation \re{3.40}, \re{3.24} and \re{3.23} by a set of equivalent equations, namely, \re{3.40}, \re{3.24} and \re{3.33} 
\begin{equation}\lae{3.33}
\begin{aligned}
g_{ij}\{\pi^{ij},\mc H\}&=(n-1) (R-2\Lam)w\sqrt g-Rw\sqrt g-(n-1)\tilde\D w \sqrt g \\
&\q\qq -\frac1{\sqrt g}G_{rs,kl}\pi^{rs}\pi^{kl}w,
\end{aligned}
\end{equation}
where $\tilde \D$ is the Laplacian with respect to the metric $g_{ij}$.
Let us formulate this claim as a theorem:
\bt
Let $N=N^{n+1}$ be a globally hyperbolic spacetime and let the metric $\bar g_{\al\bet}$ be expressed as in \re{3.2}. Then, the metric satisfies the full Einstein equations if and only if the metric is a solution of the  Hamilton equations \re{3.24} and \re{3.40} and of the equation \re{3.33}.
\et
\bp
The second Hamilton equation states
\begin{equation}
\dot\pi^{ij}=-\frac{\de\mc H}{\de g_{ij}},
\end{equation}
which is of course equal to \re{3.32}, and 
\begin{equation}\lae{4.10.1}
\begin{aligned}
-\dde{\mc H}{g_{ij}}=-\frac{\pa}{\pa g_{ij}}(\frac1{\sqrt g}G_{rs,kl}\pi^{rs}\pi^{kl})w+\dde{((R-2\Lam)w\sqrt g)}{g_{ij}}.
\end{aligned}
\end{equation}
In the lemma below we shall prove
\begin{equation}\lae{4.11.1}
\begin{aligned}
\dde{((R-2\Lam)w\sqrt g)}{g_{ij}}&=\tfrac12Rg^{ij}w\sqrt g-R^{ij}w\sqrt g\\
&\q+\{w^{ij}-\D w g^{ij}-\Lam g^{ij}w\}\sqrt g
\end{aligned}
\end{equation}
and a simple but somewhat lengthy computation will reveal
\begin{equation}\lae{4.12.1}
\begin{aligned}
-\frac{\pa}{\pa g_{ij}}(\frac1{\sqrt g}G_{rs,kl}\pi^{rs}\pi^{kl})w&= \tfrac12(\abs A^2-H^2)g^{ij}w\sqrt g\\
&\msp[-100]-2\pi^i_r\pi^{rj}w\frac1{\sqrt g} +\frac2{n-1}\pi^{ij}\pi^r_rw\frac1{\sqrt g},
\end{aligned}
\end{equation}
where the indices are lowered with the help of $g_{ij}$ and we further conclude
\begin{equation}\lae{4.12.1.1}
\begin{aligned}
\msp[-79]&-g_{ij}\frac{\pa}{\pa g_{ij}}(\frac1{\sqrt g}G_{rs,kl}\pi^{rs}\pi^{kl})w
\\&= \frac n2(\abs A^2-H^2)w\sqrt g
-2(\abs A^2-H^2)w{\sqrt g}\\
&=(\frac n2-1)(\abs A^2-H^2)w\sqrt g-\frac1{\sqrt g}G_{rs,kl}\pi^{rs}\pi^{kl}w
\end{aligned}
\end{equation}

On the other hand, the Hamilton density is equal to 
\begin{equation}\lae{4.8.1}
\mc H=-2\{\ein-\Lam\}w\sqrt g
\end{equation}
because of the Gau{\ss} equation. Hence,
\begin{equation}\lae{3.39} 
\frac12\{\abs A^2-H^2\}w\sqrt g=\frac12 (R-2\Lam)w\sqrt g
\end{equation}
iff the Hamilton constraint is valid, from which the proof of the theorem immediately follows.
\ep

\bl\lal{3.3}
Let $M$ be a Riemannian manifold with metric $g_{ij}$, scalar curvature $R$ and let $w\in C^2(M)$ and $\Lam\in\R[]$, then the equation \re{4.11.1} is valid.
\el

\bp
It suffices to consider the term 
\begin{equation}
\dde{(Rw\sqrt g)}{g_{ij}},
\end{equation}
since the result for the second term is trivial.

Let $\Om\su M$ be open and bounded and define the functional
\begin{equation}
J=\int_\Om Rw\sqrt g.
\end{equation}
Let $g_{ij}(\e)$ be a variation of $g_{ij}$ with support in $\Om$ such that
\begin{equation}
g_{ij}=g_{ij}(0)
\end{equation}
and denote differentiation with respect to $\e$ by a dot or prime, then the first variation of $J$ with respect to this variation is equal to
\begin{equation}
\begin{aligned}
\dot J(0)&=\int_\Om\{\dot g^{ij}R_{ij}+g^{ij}\dot R_{ij}\}w\sqrt g+\int_\Om Rw\sqrt g'.
\end{aligned}
\end{equation}
Again we only consider the non-trivial term
\begin{equation}\lae{4.19.1}
\int_\Om g^{ij}\dot R_{ij}w\sqrt g.
\end{equation}
It is well known that
\begin{equation}
\dot R_{ij}=-(\dot\C^k_{ik})_{;j}+(\dot\C^k_{ij})_{;k},
\end{equation}
where the semicolon indicates covariant differentiation, $\dot\C^k_{ij}$ is a tensor. Hence, we deduce that \re{4.19.1} is equal to
\begin{equation}
\int_\Om\{g^{ij}\dot\C^k_{ik}w_j-g^{ij}\dot\C^k_{ij}w_k\}\sqrt g
\end{equation}
which in turn can be expressed as
\begin{equation}
\begin{aligned}
&\int_\Om g^{ij}g^{kl}\tfrac12(\dot g_{il;k}+\dot g_{kl;i}-\dot g_{ik;l})w_j\\
&-\int_\Om g^{ij}g^{kl}\tfrac12(\dot g_{il;j}+\dot g_{jl;i}-\dot g_{ij;l})w_k,
\end{aligned}
\end{equation}
where we omitted the notation of the density $\sqrt g$. Let us agree that each row of the preceding expression contains three integrals. Then the first integrals in each row cancel each other, the second in the first row is equal to the third integral in the second row and the third integral in the first row is equal to the second integral in the second row. Therefore, we obtain by integrating by parts
\begin{equation}
\begin{aligned}
-\int_\Om\D wg^{kl}\dot g_{kl}+\int_\Om w^l_i\dot g^i_l
=\int_\Om\{-\D wg^l_i+w^l_i\}\dot g^i_l
\end{aligned}
\end{equation}
and conclude
\begin{equation}
\dde{(Rw\sqrt g)}{g_{ij}}=(\tfrac12 Rg^{ij}-R^{ij})w\sqrt g+(w^{ij}-\D wg^{ij})\sqrt g.
\end{equation}
\ep

\section{The quantization} \las{4}
For the quantization of the Hamiltonian setting we use the same approach as in our former paper \cite{cg:qgravity}, at least in the beginning: First, we replace all densities by tensors, by choosing a fixed Riemannian metric in $\so$
\begin{equation}\lae{4.1} 
\chi=(\chi_{ij}(x)),
\end{equation}
and, for a given metric $g=(g_{ij}(t,x))$, we define
\begin{equation}
\f=\f(x,g_{ij})=\big(\frac{\det g_{ij}}{\det \chi_{ij}}\big)^\frac12
\end{equation}
such that the Einstein-Hilbert functional $J$ in \fre{3.17} can be written in the form
\begin{equation}
J=\int_a^b\int_\Om\{\frac14G^{ij,kl}\dot g_{ij}\dot g_{kl}w^{-2}+(R-2\Lam)\}w\f\sqrt \chi.
\end{equation}
The Hamilton density $\mc H$ is then replaced by the function
\begin{equation}
H=\{\f^{-1}G_{ij,kl}\pi^{ij}\pi^{kl}-(R-2\Lam)\f\}w,
\end{equation}
where now
\begin{equation}
\pi^{ij}=-\f G^{ij,kl}h_{kl}
\end{equation}
and
\begin{equation}
h_{ij}=-\f^{-1}G_{ij,kl}\pi^{kl}.
\end{equation}
The effective Hamiltonian is of course
\begin{equation}
w^{-1}H.
\end{equation}
Fortunately, we can, at least locally, assume
\begin{equation}
w=1
\end{equation}
by choosing an appropriate coordinate system: Let $(t_0,x_0)\in N$ be an arbitrary point, then consider the Cauchy hypersurface
\begin{equation}
M(t_0)=\{t_0\}\times \so
\end{equation}
and look at a tubular neighbourhood of $M(t_0)$, i.e., we define new coordinates $(t,x^i)$, where $(x^i)$ are coordinate for $\so$ near $x_0$ and $t$ is the signed Lorentzian distance to $M(t_0)$ such that the points
\begin{equation}
(0,x^i)\in M(t_0).
\end{equation}
The Lorentzian metric of the ambient space then has the form
\begin{equation}
d\bar s^2=-dt^2+g_{ij} dx^idx^j.
\end{equation}

Secondly, we use the same model as in \cite[Section 3]{cg:qgravity}: The Riemannian metrics $g_{ij}(t,\cdot)$ are elements of the bundle $T^{0,2}(\so)$. Denote by $E$ the fiber bundle with base $\so$ where the fibers consists of the Riemannian metrics $(g_{ij})$. We shall consider each fiber to be a Lorentzian manifold equipped with the DeWitt metric. Each fiber $F$ has dimension
\begin{equation}
\dim F=\frac{n(n+1)}2\equiv m+1.
\end{equation}
Let $(\xi^a)$, $0\le a\le m$, be  coordinates for a local trivialization such that
\begin{equation}
g_{ij}(x,\xi^a)
\end{equation}
is a local embedding. The DeWitt metric is then expressed as
\begin{equation}
G_{ab}=G^{ij,kl}g_{ij,a}g_{kl,b},
\end{equation}
where a comma indicates partial differentiation.  The Hamiltonian is then expressed as
\begin{equation}
H=\f^{-1}G^{ab}\pi_a\pi_b-(R-2\Lam)\f,
\end{equation}
\cf \cite[equ. (3.55)]{cg:qgravity}. The fibers equipped with the metric
\begin{equation}\lae{4.16}
(\f G_{ab})
\end{equation}
are then globally hyperbolic Lorentzian manifolds. The hypersurfaces
\begin{equation}
\{\f=\const\}
\end{equation}
are Cauchy hypersurfaces.

Let $F=F(x)$ be a fiber and set
\begin{equation}
\tau=\log\f,
\end{equation}
then $\tau$ is a time function. In the Gaussian coordinate system $(\tau,\xi^A)$, $1\le A\le m$, corresponding to the hypersurface
\begin{equation}\lae{4.19}
M=\{\f=1\}=\{\tau=0\}
\end{equation}
the metric \re{4.16} has the form
\begin{equation}\lae{4.20}
ds^2=\frac{4(n-1)}n\f\{-d\tau^2+G_{AB}d\xi^Ad\xi^B\}.
\end{equation}
where the Riemannian metric $G_{AB}$ is independent of $\tau$
\begin{equation}
\pde{G_{AB}}\tau=0.
\end{equation}
When we work in a local trivialization of $E$, the coordinates $\xi^A$ are independent of $x$.
\bl\lal{4.1}
The function $\f$ is independent of $x$.
\el
\bp
Let
\begin{equation}\lae{4.22}
g_{ij}(x,\tau,\xi^A)
\end{equation}
be the local embedding in $E$, then we have
\begin{equation}
\dot g_{ij}=\pde{g_{ij}}\tau=\frac2n g_{ij},
\end{equation}
\cf \cite[equ. (4.13)]{cg:qgravity}, hence we conclude
\begin{equation}\lae{4.24}
\begin{aligned}
g_{ij}&=e^{\frac2n\tau}g_{ij}(x,0,\xi^A)\\
&\equiv e^{\frac2n\tau}\s_{ij}(x,\xi^A),
\end{aligned}
\end{equation}
where
\begin{equation}
\s_{ij}=g_{ij}(0)\in M
\end{equation}
and we further deduce
\begin{equation}\lae{4.26}
\f^2=\frac{\det g_{ij}}{\det \chi_{ij}}=e^{2\tau}\frac{\det\s_{ij}}{\det\chi_{ij}}.
\end{equation}
In the embedding \re{4.22} $\tau$ is considered to be independent of $x$ being the time component of a coordinate system satisfying \re{4.19} and \re{4.20}. Therefore, we infer from \re{4.26} 
\begin{equation}\lae{4.27} 
\det \s_{ij}=\det \chi_{ij},
\end{equation}
proving the lemma.
\ep

We can now quantize the Hamiltonian setting using the original variables $g_{ij}$ and $\pi^{ij}$. We consider the bundle $E$ equipped with the metric \re{4.20}, or equivalently,
\begin{equation}\lae{4.28}
(\f G^{ij,kl}),
\end{equation}
which is the \tit{covariant} form, in the fibers and with the Riemannian metric $\chi$ in $\so$. Furthermore, let
\begin{equation}
C^\un_c(E)
\end{equation}
be the space of real valued smooth functions with compact support in $E$.

In the quantization process, where we choose $\hbar=1$, the variables $g_{ij}$ and $\pi^{ij}$ are then replaced by operators $\hat g_{ij}$ and $\hat\pi^{ij}$ acting in $C^\un_c(E)$ satisfying the commutation relations
\begin{equation}
[\hat g_{ij},\hat\pi^{kl}]=i\de^{kl}_{ij},
\end{equation}
while all the other commutators vanish. These operators are realized by defining $\hat g_{ij}$ to be the multiplication operator
\begin{equation}
\hat g_{ij} u=g_{ij}u
\end{equation}
and $\hat \pi^{ij}$ to be the \tit{functional} differentiation
\begin{equation}
\hat\pi^{ij}=\frac1i \dde{}{g_{ij}},
\end{equation}
i.e., if $u\in C^\un_c(E)$, then
\begin{equation}
\dde{u}{g_{ij}}
\end{equation}
is the Euler-Lagrange operator of the functional
\begin{equation}
\int_{\so}u\sqrt\chi\equiv\int_\so u.
\end{equation}
Hence, if $u$ only depends on $(x,g_{ij})$ and not on derivatives of the metric, then
\begin{equation}
\dde{u}{g_{ij}}=\pde u{g_{ij}}.
\end{equation}
Therefore, the transformed Hamiltonian $\hat H$ can be looked at as the hyperbolic differential operator
\begin{equation}\lae{4.36}
\hat H=-\D-(R-2\Lam)\f,
\end{equation}
where $\D$ is the Laplacian of the metric in \re{4.28} acting on functions
\begin{equation}
u=u(x,g_{ij}).
\end{equation}
We used this approach in \cite{cg:qgravity} to transform the Hamilton constraint to the Wheeler-DeWitt equation
\begin{equation}
\hat Hu=0\qq \text{in } E
\end{equation}
which can be solved with suitable Cauchy conditions. However, the Hamiltonian in the Wheeler-DeWitt equation is a differential operator that only acts in the fibers of $E$ and not in the base space $\so$ which seems to be insufficient. This short-coming will be eliminated when, instead of the explicit Hamilton constraint, its equivalent  implicit version, equation \fre{3.33} is quantized: Following Dirac the Poisson brackets are replaced by $\frac1i$ times the commutators in the quantization process, $\hbar=1$, i.e., we obtain
\begin{equation}
\{\pi^{ij},H\}\ra i[\hat H, \hat\pi^{ij}].
\end{equation}
Dropping the hats in the following to improve the readability equation \re{3.33} is transformed to
\begin{equation}\lae{4.40}
\begin{aligned}
ig_{ij}[H,\pi^{ij}]&=(n-1) (R-2\Lam)\f-R  \f+\D,
\end{aligned}
\end{equation}
where $\D$ is the Laplace operator with respect to the fiber metric.

Now, we have
\begin{equation}\lae{4.41}
\begin{aligned}
i[H,\pi^{ij}]&=[H,\dde{}{g_{ij}}]\\
&= [-\D,\dde{}{g_{ij}}]-[(R-2\Lam)\f,\dde{}{g_{ij}}],
\end{aligned}
\end{equation}
\cf \re{4.36}. Since we apply both sides to functions $u\in C^\un_c(E)$ 
\begin{equation}\lae{4.42}
[-\D,\dde{}{g_{ij}}]u=[-\D,\pde{}{g_{ij}}]u=-R^{ij}_{\hp{ij},kl}u^{kl},
\end{equation}
because of the Ricci identities, where
\begin{equation}
R^{ij}_{\hp{ij},kl}
\end{equation}
is the Ricci tensor of the fiber metric \re{4.28} and
\begin{equation}
u^{kl}=\pde u{g_{kl}}
\end{equation}
is the gradient of $u$.

For the second commutator on the right-hand side of \re{4.41} we obtain
\begin{equation}\lae{4.45}
\begin{aligned}
-[(R-2\Lam)\f,\dde{}{g_{ij}}]u=-(R-2\Lam)\f\pde u{g_{ij}}+\dde{}{g_{ij}}\{(R-2\Lam)u\f\},
\end{aligned}
\end{equation}
where the last term is the Euler-Lagrange operator of the functional
\begin{equation}
\begin{aligned}
\int_\so (R-2\Lam)u\f&\equiv \int_\so(R-2\Lam)u\f\sqrt\chi\\
&=\int_\so(R-2\Lam)u\sqrt g
\end{aligned}
\end{equation}
with respect to the variable $g_{ij}$, since the scalar curvature $R$ depends on the derivatives of $g_{ij}$. From \re{4.11.1} and the proof of \frl{3.3} we infer  
\begin{equation}\lae{4.47} 
\begin{aligned}
\dde{}{g_{ij}}\{(R-2\Lam)u\f\}&=\frac12 (R-2\Lam) g^{ij}u\f-R^{ij}u\f\\
&+\f\{u_{;}^{\hp;ij}-\tilde\D u g^{ij}\}+(R-2\Lam)\f\pde u{g_{ij}},
\end{aligned}
\end{equation}
where the semicolon indicates covariant differentiation in $\so$ with respect to the metric $g_{ij}$, $\tilde \D$ is the corresponding Laplacian, and where we observe that the fundamental lemma of the calculus of variations has been applied to functions in $L^2(\so,\sqrt\chi)$, i.e.,
\begin{equation}
\int_\so f\h\sqrt g=\int_\so f\h\f\sqrt\chi;
\end{equation}
here we have 
\begin{equation}
f\in C^0(\so),\q \h\in C^\un_c(\so).
\end{equation}
We also note that
\begin{equation}
\begin{aligned}
D_ku&= \pde u{x^k}+\pde u{g_{ij}}\pde{g_{ij}}{x^k}\\
&=\pde u{x^k}
\end{aligned}
\end{equation}
in Riemannian normal coordinates.

Hence, we conclude that equation \re{4.40} is equivalent to
\begin{equation}\lae{4.59}
\begin{aligned}
-\D u -(n-1)\f\tilde \D u-\frac {n-2}2\f(R-2\Lam)u=0
\end{aligned}
\end{equation}
in $E$, since
\begin{equation}
g_{ij}R^{ij}_{\hp{ij},kl}=0
\end{equation}
for
\begin{equation}\lae{4.61}
\frac1{\sqrt{n(n-1)\f}}g_{ij}
\end{equation}
is the future directed unit normal of the Cauchy hypersurfaces $\{\f=\const\}$: The gradient of $\f$
\begin{equation}
\pde\f{g_{ij}}=\frac12\f g^{ij}
\end{equation}
is a past directed normal in \tit{covariant} notation. Its contravariant version has the form
\begin{equation}
\f^{-1}G_{ij,kl}g^{kl}\frac12\f=-\frac1{2(n-1)}g_{ij}.
\end{equation}
Therefore, the vector in \re{4.61} is future directed and has unit length as can easily be checked.

Now, let us choose a coordinate system $(\tau,\xi^A)$ associated with the Cauchy hypersurface
\begin{equation}
M=\{\f=1\}
\end{equation}
and express the metric as in \re{4.20}. The time coordinate $\tau$ is defined as
\begin{equation}
\tau=\log\f.
\end{equation}
Let $t$ be the time function
\begin{equation}\lae{4.58}
t=\sqrt\f=e^{\frac12\tau},
\end{equation}
then
\begin{equation}
dt^2=\frac14\f d\tau^2
\end{equation}
and we conclude that the fiber metric can be expressed as
\begin{equation}\lae{4.68}
ds^2=-\frac{16(n-1)}ndt^2+\frac{4(n-1)}nt^2 G_{AB}d\xi^Ad\xi^B,
\end{equation}
where $G_{AB}$ is independent of $t$. We also emphasize that $t$ is independent of $x$, \cf \rl{4.1}. 

Let $(\xi^a)=(t,\xi^A)$, $0\le a\le m$, be the coordinates such that
\begin{equation}
\xi^0=t\q\wed\q 1\le A\le m,
\end{equation}
then we immediately deduce from \re{4.68} or \re{4.20} that the Ricci tensor satisfies
\begin{equation}
R_{0a}=0\qq\A\, 0\le a\le m.
\end{equation}
Since the determinant of the metric in \re{4.68} is equal to
\begin{equation}
\abs{\det(G_{ab})}=16(\tfrac {n-1}n)\{4(\tfrac {n-1}n)\}^mt^{2m}\det(G_{AB})
\end{equation}
 we conclude that the equation \re{4.59} can be expressed in the form
\begin{equation}\lae{4.74}
\begin{aligned}
&\frac1{16}\frac{n}{n-1}t^{-m}\pde{(t^m\dot u)}t -\frac14 \frac n{n-1}t^{-2}\D_Gu\\
&\qq\qq-(n-1) t^2 \tilde\D u-\frac {n-2}2t^2(R-2\Lam)u=0,
\end{aligned}
\end{equation}
where $\D_G$ is the Laplacian with respect to the metric $G_{AB}$.

For any point
\begin{equation}
(x,g_{ij})\in E
\end{equation}
the metric can be written in the form
\begin{equation}
g_{ij}=t^\frac4n\s_{ij},
\end{equation}
where $\s_{ij}$ is independent of $t$ and 
\begin{equation}
\det \s_{ij}=\det \chi_{ij},
\end{equation}
\cf \re{4.24} and \re{4.27}. Hence, we can write
\begin{equation}
\tilde \D u =t^{-\frac4n}\tilde\D_{\s_{ij}}u.
\end{equation}
Thus, equipping $E$ with the metric 
\begin{equation}\lae{4.79}
\begin{aligned}
d\bar s^2&=-\frac{16(n-1)}{n}dt^2+\frac{4(n-1)}nt^2 G_{AB}d\xi^Ad\xi^B+\frac1{n-1}\s_{ij}dx^idx^j\\
&\equiv G_{ab}d\xi^ad\xi^b+\frac1{n-1}\s_{ij}dx^idx^j\\
&\equiv G_{\al\bet}d\zeta ^\al d\zeta^\bet,
\end{aligned}
\end{equation}
where $0\le a\le m$ and $\xi^0=t$. We call $G_{ab}$ the fiber metric and $\s_{ij}$ the base metric, which are to be evaluated at the points 
\begin{equation}
(x,\xi^a)\equiv (x,g_{ij})=(x,t^\frac4n\s_{ij}).
\end{equation}
Beware that
\begin{equation}
\s_{ij}=\s_{ij}(x,\xi^A)\in E_1,
\end{equation}
where $E_1$ is the subbundle
\begin{equation}
E_1=\{t=1\}.
\end{equation}
 This metric the operator $P$ in \re{4.74} is a symmetric hyperbolic differential operator
\begin{equation}
Pu=-D_\al(a^{\al\bet}D_\bet u),
\end{equation}
where the derivatives are covariant derivatives with respect to the metric in \re{4.79} and the coefficients $a^{\al\bet}$ represent a Lorentzian metric. However, it is not  normally hyperbolic, i.e., its main part is not identical with the Laplacian of the ambient metric. Nevertheless, we can consider $P$ as a normally hyperbolic operator by equipping $E$ with the metric
\begin{equation}
\begin{aligned}
d\tilde s^2&=-\frac{16(n-1)}{n}dt^2+\frac{4(n-1)}nt^2 G_{AB}d\xi^Ad\xi^B\\
&\msp[140]+\frac1{n-1}t^{\frac4n-2}\s_{ij}dx^idx^j\\
&\equiv \tilde G_{\al\bet}d\zeta^\al d\zeta^\bet,
\end{aligned}
\end{equation}
though, of course, $P$ is not symmetric in this metric. 

Let $E$, $\tilde E$ be the bundles
\begin{equation}
(E,G_{\al\bet})\q\wed\q (E,\tilde G_{\al\bet})
\end{equation}
respectively, and $E_1$ \resp $\tilde E_1$ the corresponding subbundles defined by
\begin{equation}
\{t=1\}.
\end{equation}
We shall now prove that $E$ and $\tilde E$ are both globally hyperbolic manifolds and the subbundles $E_1$ \resp $\tilde E_1$, or more generally, the subbundles $E_1(\tau)$ \resp $\tilde E_1(\tau)$, defined by
\begin{equation}
\{t=\tau\},\q\tau>0,
\end{equation}
Cauchy hypersurfaces provided the base space $\so$ is either compact or a homogeneous space for a suitable metric $\rho_{ij}$.

\bl\lal{4.2}
The bundles $E$ and $\tilde E$ are both globally hyperbolic manifolds, if $\so$ is either compact or a homogeneous space for a suitable metric $\rho_{ij}$, and the hypersurfaces $E_1(\tau)$ \resp $\tilde E_1(\tau)$ are Cauchy hypersurfaces.
\el
\bp
We shall only prove that $E$ is globally hyperbolic, since the proof for $\tilde E$ is essentially identical. We shall show that $E_1$ is a Cauchy hypersurface. The arguments will then also apply in case of the hypersurfaces $E_1(\tau)$. The proof will be similar to the proof of \cite[Lemma 4.3]{cg:qgravity}, where we proved that the fibers of $E$ are globally hyperbolic. The fact that we now consider  the whole bundle creates a small complication which will be handled by the additional assumption on $\so$.

We shall now prove that $E_1$ is a Cauchy hypersurface implying that $E$ is globally hyperbolic. Let us argue by contradiction. Thus, let 
\begin{equation}
\ga(s)=(\ga^\al(s)),\q s\in I=(a,b),
\end{equation}
be an inextendible future directed causal curve in $E$ and assume that $\ga$ does not intersect $E_1$. We shall show that this will lead to a contradiction. It is also obvious that $\ga$ can meet $E_1$ at most once.

Assume that there exists $s_0\in I$ such that
\begin{equation}
t(\ga(s_0))<1
\end{equation}
and assume from now on that $s_0$ is the left end point of $I$. Since $t$ is continuous, the whole curve $\ga$ must be contained in the past of $E_1$.

$\ga$ is causal, i.e.,
\begin{equation}
\begin{aligned}
\frac1{n-1}\s_{ij}\dot x^i\dot x_j+\frac{4(n-1)}nt^2G_{AB}\dot\ga^A\dot\ga^B\le \frac{16(n-1)}{n}\abs{\dot\ga^0}^2
\end{aligned}
\end{equation}
and thus
\begin{equation}
\sqrt{\frac1{n-1}\s_{ij}\dot x^i\dot x_j+\frac{4(n-1)}nt^2G_{AB}\dot\ga^A\dot\ga^B}\le 4\dot\ga^0,
\end{equation}
since $\ga$ is future directed.

Let
\begin{equation}
\tilde\ga=(x^i,\ga^A)
\end{equation}
be the projection of $\ga$ onto $E_1$, then the length of $\tilde\ga$ is bounded
\begin{equation}
\begin{aligned}
L(\tilde\ga)&\le \int_I\sqrt{\frac1{n-1}\s_{ij}\dot x^i\dot x_j+\frac{4(n-1)}nG_{AB}\dot\ga^A\dot\ga^B}\\
&\le 4(1-t(s_0))<4.
\end{aligned}
\end{equation}
Expressing the quadratic form
\begin{equation}
G_{AB}\dot\ga^A\dot\ga^B
\end{equation}
in $E_1$ in the coordinates $(g_{ij})=(\s_{ij})$, we have
\begin{equation}
\begin{aligned}
G_{AB}\dot\ga^A\dot\ga^B&=\s^{ik}\s^{jl}\dot\s_{ij}\dot\s_{kl}\\
&\equiv\norm{\dot\s_{ij}}^2,
\end{aligned}
\end{equation}
since the right-hand side is exactly
\begin{equation}
G^{ij,kl}\dot\s_{ij}\dot\s_{kl},
\end{equation}
if
\begin{equation}
\dot\s_{ij}\in T(E_1).
\end{equation}
Hence, we infer, in view of \cite[Lemma 14.2]{hamilton:ricci}, that the metrics $(\s_{ij}(s))$ are all uniformly equivalent in $I$ and converge to a positive definite metric when $s$ tends to $b$. It remains to prove that the points $(x^i(s))$ are precompact in $\so$, then we would have derived a contradiction.

If $\so$ is compact then the precompactness of $(x^i(s))$ is trivial, thus let us assume that $(\so, \rho_{ij})$ is a homogeneous space. Then $\s_{ij}(s_0)$ is equivalent to $\rho_{ij}(x(s_0))$ and hence, in view of the homogeneity, $\s_{ij}(s)$ is uniformly equivalent to $\rho_{ij}(x(s))$ for all $s\in I$, and we conclude
\begin{equation}
\int_I\sqrt{\rho_{ij}\dot x^i\dot x^j}\le \const
\end{equation}
proving the precompactness. $E_1$ is therefore a Cauchy hypersurface and $E$ is globally hyperbolic.
\ep

\br
Since $\tilde E$ is globally hyperbolic and $P$ is a normally hyperbolic differential operator the Cauchy problems
\begin{equation}\lae{4.100}
\begin{aligned}
Pu&=f,\\
\fv u{\tilde E_1(\tau)}&=u_0,\\
\fv {u_\al\tilde\nu^\al}{\tilde E_1(\tau)}&=u_1
\end{aligned}
\end{equation}
have unique solutions
\begin{equation}
u\in C^\un(\tilde E)
\end{equation}
for given values $u_0,u_1\in C^\un_c(\tilde E_1(\tau))$ and $f\in C^\un_c(\tilde E)$ such that
\begin{equation}
\supp u\su J^{\tilde E}(K),
\end{equation}
where
\begin{equation}
K=\supp u_0\uu \supp u_1\uu\supp f,
\end{equation}
\cf \cite{gunther:book, baer:book, ginoux:fredenhagen}.
\er

Since $E$, $\tilde E$ and $E_1(\tau)$ \resp $\tilde E_1(\tau)$ coincide as sets and the normals $(\nu^\al)$ \resp $\tilde\nu^\al)$ are also identical
\begin{equation}
\tilde\nu=\nu
\end{equation}
we immediately deduce that the Cauchy problems \re{4.100} are also uniquely solvable in $E$. Using this information we then could derive the existence of the fundamental solutions $F_\pm$ for $P$ in $E$ and also the existence of the advanced \resp retarded Green's operators $G_\pm$ of $P$, \cf \cite[Theorem 4]{ginoux:fredenhagen}.

However, we would like to show how the fundamental solutions $\tilde F_\pm$ of $P$ in $\tilde E$ can easily be transformed to yield fundamental solutions of $P$ in $E$ and similarly  the Green's functions $\tilde G_\pm$. This process is valid in general pseudo-riemannian manifolds, and thus also valid for elliptic operators, however, we shall only consider Lorentzian manifolds. The notations $N$ \resp $\tilde N$ refer to the same manifold $N$ equipped with the metrics $g_{\al\bet}$ \resp $\tilde g_{\al\bet}$.

\bd
Let $T\in \mc D'(N)$ be a distribution and let $\sqrt {\abs g}$ be the volume element in $N$, where
\begin{equation}
g=\det g_{\al\bet},
\end{equation}
then we use the notation
\begin{equation}
\spd T{\h\sqrt{\abs g}}
\end{equation}
or
\begin{equation}
T[\h\sqrt{\abs g}]
\end{equation}
to refer to \cq{T acts on $\h$} instead of the usual symbols
\begin{equation}
\spd T\h
\end{equation}
or
\begin{equation}
T[\h].
\end{equation}
If $P$ is a differential operator in $N$ and $P^*$ its formal adjoint, then
\begin{equation}
\spd{PT}{\h\sqrt{\abs g}}=\spd T{(P^*\h)\sqrt{\abs g}}.
\end{equation}
\ed
We found this notation in \cite[Definition 2.8.1, p. 60]{friedlander:book}

\bl
Let $T\in \mc D'(N,\tilde g)$ and let $g$ be a another smooth metric in $N$ and set
\begin{equation}
\psi=\frac{\sqrt{\abs{\tilde g}}}{\sqrt{\abs g}},
\end{equation}
then
\begin{equation}
\psi T\in \mc D'(N,g)
\end{equation}
and
\begin{equation}
\spd{\psi T}{\h\sqrt{\abs g}}=\spd T{\h\sqrt{\abs{\tilde g}}}\qq\A\,\h\in C^\un_c(N).
\end{equation}
\el
\bp
Follows immediately from the definition of $\psi T$
\begin{equation}
\spd{\psi T}{\h\sqrt{\abs g}}=\spd T{\psi\h\sqrt{\abs g}}=\spd T{\h\sqrt{\abs{\tilde g}}}.
\end{equation}
\ep

As an application we obtain:
\bc\lac{4.6}
Let $\tilde F_\pm$ \resp $\tilde G_\pm$ be the fundamental solutions of $P$ in $\tilde E$ \resp the advanced and retarded Green's operators, and define
\begin{equation}
\psi=\frac{\sqrt{\abs{\tilde G}}}{\sqrt{\abs G}}=t^{2-n},
\end{equation}
then
\begin{equation}\lae{4.116}
F_\pm=\psi\tilde F_\pm
\end{equation}
are fundamental solutions of $P$ in $E$ and
\begin{equation}\lae{4.117}
G_\pm=\psi\tilde G_\pm
\end{equation}
the advanced and retarded Green's operators.
\ec
\bp
\cq{\re{4.116}}\q We have
\begin{equation}\lae{4.118}
\begin{aligned}
F_\pm[\h\sqrt{\abs G}\,]&=\psi\tilde F_\pm[\h\sqrt{\abs G}\,]\\
&=\tilde F_\pm[\h\sqrt{\abs{\tilde G}}\,]
\end{aligned}
\end{equation}
and
\begin{equation}
PF_\pm[\h\sqrt{\abs G}\,]=P\tilde F_\pm[\h\sqrt{\abs{\tilde G}}\,]=\h.
\end{equation}

\cvm
\cq{\re{4.117}}\q To prove the second claim we note that the Green's operators are defined as maps
\begin{equation}
C^\un_c(E)\ra C^\un(E)
\end{equation}
by the definition
\begin{equation}
G_\pm[\h\sqrt{\abs G}\,](p)=F_\pm(p)[\h\sqrt{\abs G}\,],\qq p\in E.
\end{equation}
Now, from \re{4.118} we deduce
\begin{equation}
\begin{aligned}
F_\pm(p)[\h\sqrt{\abs G}\,]&=\tilde F_\pm(p)[\h\sqrt{\abs{\tilde G}}\,]\\
&=\tilde G_\pm[\h\sqrt{\abs{\tilde G}}\,](p)\\
&=\psi\tilde G_\pm[\h\sqrt{\abs G}\,](p).
\end{aligned}
\end{equation}
\ep

\br\lar{4.7}
Let $G$ be the Green's operator of $P$ in $E$
\begin{equation}
G=G_+-G_-,
\end{equation}
then
\begin{equation}
N(P)=\{Gu:u\in C^\un_c(E)\}
\end{equation}
is the kernel of $P$. Its elements are smooth functions which are spacelike compact; however, this condition is  strictly correct only in $\tilde E$, since the light cones in $\tilde E$ and $E$ are different. Fortunately, we only need one special property of spacelike compact functions, namely, that their restrictions to Cauchy hypersurfaces have compact support. This will be case in $E$, if we only consider the Cauchy hypersurfaces $E_1(\tau)$, as we shall prove in the lemma below.
\er

\bl
The compact subsets of $\tilde E_1(\tau)$ are also compact in $E_1(\tau)$ and vice versa.
\el
\bp
The Cauchy hypersurfaces $E_1(\tau)$ \resp $\tilde E_1(\tau)$ carry the same topology, since their induced metrics are uniformly equivalent as one easily checks.
\ep

\section{The second quantization}\las{5}
Let us first summarize some facts about the Green's operators $G_\pm$ of $P$ in $E$ which are still valid even though $P$ is not normally hyperbolic.
\bl
Let $G_\pm$ \resp $\tilde G_\pm$ be the Green's operators of $P$ in $E$ \resp $\tilde E$, then
\begin{equation}\lae{5.1}
G_\pm:C^\un_c(E)\ra C^\un(E)
\end{equation}
\begin{equation}\lae{5.2}
P\circ G_\pm=\fv{G_\pm\circ P}{C^\un_c(E)}=\fv{\id}{C^\un_c(E)}
\end{equation}
\begin{equation}\lae{5.3}
\supp(G_\pm u)=\supp(\tilde G_\pm u)\q\A\,u\in C^\un_c(E)
\end{equation}
\begin{equation}\lae{5.4}
\supp G_+u\su J_+^{\tilde  E}(\supp u)\q \A\,u\in C^\un_c(E)
\end{equation}
\begin{equation}\lae{5.5}
\supp G_-u\su J_-^{\tilde  E}(\supp u)\q \A\,u\in C^\un_c(E)
\end{equation}
\begin{equation}\lae{5.6}
\supp G_+u\ii\supp G_-v \q \text{is compact}
\end{equation}
for all $u,v\in C^\un_c(E)$, and
\begin{equation}\lae{5.7} 
G_\pm^*=G_\mp.
\end{equation}
\el
\bp
The properties \re{5.1} and \re{5.2} immediately follow from the corresponding relations for $\tilde G\pm$ of $P$ in $\tilde E$ and the fact that
\begin{equation}
G_\pm=t^{2-n}\tilde G_\pm,
\end{equation}
\cf \frc{4.6}. The preceding relation also proves the properties \re{5.3}, \re{5.4}, \re{5.5} and \re{5.6}, since the topologies of $E$ and $\tilde E$ are identical. 

It remains to prove \re{5.7}. Let $u,v\in C^\un_c(E)$, then
\begin{equation}
\begin{aligned}
\int_E\spd{G_\pm u}v&=\int_E\spd{G_\pm u}{PG_\mp v}\\ 
&= \int_E\spd{PG_\pm u}{G_\mp v}\\
&=\int_E \spd{u}{G_\mp v},
\end{aligned}
\end{equation}
where the partial integration is justified because of \re{5.6}, and the scalar product is just normal multiplication.
\ep
\bl
Let $E_1(\tau)$ be one of the special Cauchy hypersurfaces in $E$, then
\begin{equation}\lae{5.10}
\begin{aligned}
\int_E\spd{u}{Gv}=\int_{E_1(\tau)}\{\spd{D_\nu(Gu)}{Gv}-\spd{Gu}{D_\nu Gv}\},
\end{aligned}
\end{equation}
for all $u,v\in C^\un_c(E)$, where $\nu$ is the future directed normal of $E_1(\tau)$.
\el
\bp
Let $E_+$, $E_-$ be defined by
\begin{equation}
E_+=\{t>\tau\}
\end{equation}
and
\begin{equation}
E_-=\{t<\tau\},
\end{equation}
then
\begin{equation}
\int_E\spd u{Gv}=\int_{E_+}\spd u{Gv}+\int_{E_-}\spd u{Gv}.
\end{equation}
Now, in $E_+$ we have
\begin{equation}
PG_-u=u
\end{equation}
and
\begin{equation}
PGv=0=GPv.
\end{equation}
Moreover, 
\begin{equation}
\supp(G_- u)\ii E_+\q\text{is compact},
\end{equation}
since
\begin{equation}
\supp(\tilde G_- u)\ii \tilde E_+\q\text{is compact},
\end{equation}
hence we obtain by partial integration
\begin{equation}
\begin{aligned}
\int_{E_+}\spd{PG_-u}{Gv}=-\int_{E_1(\tau)}\spd{D_\nu G_-u}{Gv}+\int_{E_1(\tau)}\spd{G_-u}{D_\nu Gv}.
\end{aligned}
\end{equation}
A similar argument applies to  $E_-$ by looking at
\begin{equation}
PG_+u=0
\end{equation}
leading to 
\begin{equation}
\begin{aligned}
\int_{E_-}\spd{PG_+u}{Gv}=\int_{E_1(\tau)}\spd{D_\nu G_+u}{Gv}-\int_{E_1(\tau)}\spd{G_+u}{D_\nu Gv}.
\end{aligned}
\end{equation}
Adding these two equations implies the result.
\ep

We shall now construct a CCR representation or a Weyl system for $P$ and its kernel
\begin{equation}
N(P)=\{u\in C^\un(E):Pu=0\}=\{Gu:u\in C^\un_c(E)\}.
\end{equation}
This characterization of $N(P)$ is correct, since it is valid in $\tilde E$ and because of
\begin{equation}
PG[u\sqrt{\abs G}\,]=P\tilde G[u\sqrt{\abs{\tilde G}}\,],
\end{equation}
\cf \fre{4.117}.

There are two ways to construct a Weyl system given a formally self-adjoint, normally hyperbolic operator in a globally hyperbolic spacetime which are also applicable in our case, though $P$ is not normally hyperbolic. One possibility is to define a symplectic vector space
\begin{equation}\lae{5.23}
V=C^\un_c(e)/N(G),
\end{equation}
where $G$ is the Green's operator of $P$
\begin{equation}
G=G_+-G_-.
\end{equation}
Since
\begin{equation}\lae{5.25}
G^*=-G
\end{equation}
the bilinear form
\begin{equation}\lae{5.26}
\om(u,v)=\int_E\spd{u}{Gv},\qq u,v\in V,
\end{equation}
is skew-symmetric, non-degenerate by definition, and hence symplectic. Then, there is a canonical way to construct a corresponding Weyl system.

The second method is to pick a Cauchy hypersurface $E_1$ in $E$ and then define a quantum field $\F$ with values in the space of essentially self-adjoint operators in a corresponding symmetric Fock space.

We pick a Cauchy hypersurface $E_1=E_1(\tau)$ in $E$ and define the complex Hilbert space
\begin{equation}
H_{E_1}=L^2(E_1)\otimes \Cc=L^2(E_1,\Cc)
\end{equation}
the complexification of the real Hilbert space $L^2(E_1)$ with complexified scalar product
\begin{equation}
\spd uv_{E_1}=\int_{E_1}\spd uv_{\Cc}.
\end{equation}
We denote the symmetric Fock space of $H_{E_1}$ by $\mc F(H_{E_1})$. Let $\Theta$ be the corresponding Segal field. Since $G^*=-G$, we deduce from \re{5.4}, \re{5.6} and \frr{4.7} that
\begin{equation}
\fv{G^*u}{E_1}\in C^\un_c(E_1)\su H_{E_1}\q\A\, u\in C^\un_c(E).
\end{equation}
We can therefore define
\begin{equation}\lae{5.30}
\F_{E_1}(u)=\Theta(i\fv{(G^*u)}{E_1}-\fv{D_\nu(G^*u)}{E_1}).
\end{equation}
From the proof of \cite[Lemma 4.6.8]{baer:book} we conclude that the right-hand side of \re{5.30} is an essentially self-adjoint operator in $\mc F(H_{E_1})$.  We therefore call the map $\F_{E_1}$ from $C^\un_c(E)$ to the set of self-adjoint operators in $\mc F(H_{E_1})$ a quantum field defined in $E_1$.
\bl
The quantum field $\F_{E_1}$ satisfies the equation
\begin{equation}
P\F_{E_1}=0
\end{equation}
in the distributional sense, i.e.,
\begin{equation}
\begin{aligned}
\spd{P\F_{E_1}}u&=\spd{\F_{E_1}}{Pu}\\
&=\F_{E_1}(Pu)=0\qq\A\,u\in C^\un_c(E).
\end{aligned}
\end{equation}
\el
\bp
In view of \re{5.25} we have
\begin{equation}
G^*(Pu)=0.
\end{equation}
\ep
With the help of the quantum field $\F_{E_1}$ we shall construct a Weyl system and hence a CCR representation of the symplectic vector space $(V,\om)$ which we defined in \re{5.23} and \re{5.26}.

From \re{5.30} we conclude  the commutator relation
\begin{equation}
[\F_{E_1}(u),\F_{E_1}(v)]=i\imc\spd{iG^*u-D_\nu(G^*u)}{iG^*v-D_\nu(G^*v)}_{E_1}I,
\end{equation}
for all $u,v\in C^\un_c(E)$, \cf \cite[Proposition 5.2.3]{robinson:book2}, where both sides are defined in the algebraic Fock space $\mc F_{\tup{alg}}(H_{E_1})$. 

On the other hand   
\begin{equation}\lae{5.35}
\begin{aligned}
\imc\spd{iG^*u-D_\nu(G^*u)}{iG^*v-D_\nu(G^*v)}_{E_1}&\\
&\msp[-350]=-\imc\spd{iG^*u}{D_\nu(G^*v)}_{E_1}-\imc\spd{D_\nu(G^*u)}{iG^*v}_{E_1}\\
&\msp[-350]=\int_{E_1}\{\spd{G^*u}{D_\nu(G^*v)}-\spd{D_\nu(G^*u)}{G^*v}\}\\
&\msp[-350]=\int_{E}\spd u{Gv}
\end{aligned}
\end{equation}
in view of \re{5.10} and \re{5.25}.

As a corollary we conclude
\begin{equation}\lae{5.36}
[\F_{E_1}(u),\F_{E_1}(v)]=i\int_{E_1}\spd u{Gv}I\q\A\, u,v\in C^\un_c(E).
\end{equation}

From \cite[Proposition 5.2.3]{robinson:book2} and \re{5.35} we immediately infer
\bt\lat{5.6}
Let $(V,\om)$ be the symplectic vector space in \re{5.23} and \re{5.26} and denote by $[u]$ the equivalence classes in $V$, then
\begin{equation}
W([u])=e^{i\F_{E_1}(u)}
\end{equation}
defines a Weyl system for $(V,\om)$, where $\F_{E_1}(u)$ is now supposed to be the closure of $\F_{E_1}(u)$ in $\mc F(H_{E_1})$, i.e., $\F_{E_1}(u)$ is a self-adjoint operator. The Weyl system generates a $C^*$-algebra with unit which we call a CCR representation of $(V,\om)$.
\et
\br\lar{5.7}
Since all CCR representations of $(V,\om)$ are $^*$-isomorphic, where the isomorphism maps Weyl systems to Weyl systems, \cf \cite[Theorem 5.2.8]{robinson:book2},  this especially applies  to the CCR representations corresponding to different Cauchy hypersurfaces ${E_1=E_1(\tau)}$ and ${E_1}'=E_1(\tau')$, i.e., there exists a $^*$-isomorphism $T$ such that
\begin{equation}
T(e^{i\F_{E_1}(u)})=e^{i\F_{{E_1}'}(u)}\qq \A\, [u]\in V.
\end{equation}
\er
\section{The gravitational waves  model}\las{6}
In the previous sections we saw that the quantization of the Hamilton constraint does not yield a unique result but depends on the equation by which the Hamilton constraint is expressed. In \cite{cg:qgravity} we obtain the Wheeler-DeWitt equation  after quantization and in the previous sections the equation \fre{4.74} which differs significantly. In this section we shall propose yet another model by replacing any occurrence of the term
\begin{equation}\lae{6.1.1}
\abs A^2-H^2
\end{equation}
by
\begin{equation}
(R-2\Lam).
\end{equation}
However, when we do this on the right-hand side of \fre{3.33}, then after quantization, we would obtain an elliptic equation instead of an hyperbolic equation, namely,
\begin{equation}\lae{6.3.1} 
-(n-1)\tilde{\D} u +\frac{n-4}2(R-2\Lam)u=0 
\end{equation}
valid in $E$, which, for fixed $(t,g_{ij})$, can be looked at as an eigenvalue equation, where $\Lam$ would be a constant multiple of the eigenvalue provided $n\not=4$. In case $\so$ is compact, a spectral resolution of equation \re{6.3.1} would be possible.

However, we believe that a hyperbolic and not an elliptic equation should define the possible states of quantum gravity. In order to obtain a hyperbolic equation while eliminating any occurrences of the term in \re{6.1.1} we have to express the Hamilton constraint by a different equation. In \rs{3} the Hamilton equations only yielded the tangential Einstein equations \fre{3.22}, or equivalently,
\begin{equation}\lae{6.4.1}
\bar R_{ij}-\frac12\bar Rg_{ij}+\Lam g_{ij}=0.
\end{equation}
The Hamilton constraint expresses the normal component of the Einstein equations, where the terms \tit{tangential} und \tit{normal} refer to the foliation $M(t)$ of the spacetime $N$. This foliation is also the solution set of the geometric flow equation
\begin{equation}
\dot x=-w\nu
\end{equation}
with initial hypersurface 
\begin{equation}
M_0=\so,
\end{equation}
where $\nu$ is the past directed normal $\nu$ of the solution hypersurfaces $M(t)$, \cf \cite[equ. (2.3.25)]{cg:cp}. We shall use the evolution equation of the mean curvature $H(t)$ of the $M(t)$ to define the Hamilton constraint.  

The mean curvature satisfies the evolution equation
\begin{equation}\lae{6.7.1} 
\dot H= -\tilde\D w+\{\abs A^2+\bar R_{\al\bet}\nu^\al\nu^\bet\}w,
\end{equation}
where we embellished the Laplacian with a tilde, \cf \cite[equ. (2.3.27)]{cg:cp} observing that in that reference
\begin{equation}
e^\psi =w.
\end{equation}
To exploit this evolution equation we need the following lemma: 
\bl
Assume that the equation \re{6.4.1} is valid, then 
\begin{equation}\lae{3.30.6}
\tfrac12\bar R=\tfrac1{n-1}\{G_{\al\bet}\nu^\al\nu^\bet-\Lam\}+\tfrac{n+1}{n-1}\Lam
\end{equation}
and 
\begin{equation}\lae{3.31.6}
\bar R_{\al\bet}\nu^\al\nu^\bet=\tfrac{n-2}{n-1}\{G_{\al\bet}\nu^\al\nu^\bet-\Lam\}-\tfrac2{n-1}\Lam.
\end{equation}
\el
\bp
\cq{\re{3.30.6}}\q There holds
\begin{equation}
\bar R=g^{ij}\bar R_{ij}-\bar R_{\al\bet}\nu^\al\nu^\bet
\end{equation}
and hence
\begin{equation}\lae{3.33.6}
\bar R_{\al\bet}\nu^\al\nu^\bet +\tfrac 12\bar R=\tfrac{n-1}2\bar R-n\Lam
\end{equation}
or, equivalently,
\begin{equation}\lae{3.34.6}
\tfrac1{n-1}\{\ein-\Lam\}=\tfrac12\bar R-\tfrac{n+1}{n-1}\Lam.
\end{equation}
\cq{\re{3.31.6}}\q Combining  \re{3.33.6} and \re{3.34.6} we deduce
\begin{equation}
\rnn=\tfrac{n-2}{n-1}\{\ein-\Lam\}-\tfrac2{n-1}\Lam.
\end{equation}
\ep
We note that
\begin{equation}\lae{6.15.1} 
\pi^{ij}=(Hg^{ij}-h^{ij})\f,
\end{equation}
where $(h^{ij})$ is the contravariant version of the second fundamental form and where we also point out that, as before, we introduced the function $\f$ to replace the density $\sqrt g$ in order to deal with tensors instead of densities.

Hence, we have
\begin{equation}\lae{6.16.1}
(n-1) H\f=g_{ij}\pi^{ij}
\end{equation}
and we shall use the evolution equation of
\begin{equation}
(n-1)H\f^\frac12
\end{equation}
to express the Hamilton constraint. 
 
We immediately deduce
\begin{equation}
\begin{aligned}
(\f^\frac12)'&=\frac14 \f^\frac12 g^{ij}\dot g_{ij}\\
&=-\frac12 \f^\frac12 Hw
\end{aligned}
\end{equation}
\cf \fre{3.8}, and obtain, in view of \re{6.7.1} and \re{3.31.6}, 
\begin{equation}
\begin{aligned}
(n-1)(H\f^\frac12)'&=-(n-1)\tilde\D w \f^\frac12\\
&\q+(n-1)\{\abs A^2+\bar R_{\al\bet}\nu^\al\nu^\bet\}w \f^\frac12-\frac{n-1}2H^2\f^\frac12 w\\
&=-(n-1)\tilde\D w \f^\frac12+(n-1)(\abs A^2-H^2)\f^\frac12w\\
&\q+\frac{n-1}2H^2\f^\frac12 w+(n-2)\{\ein-\Lam\}\f^\frac12w\\
&\q -2\Lam \f^\frac12 w.
\end{aligned}
\end{equation}

Employing now the Hamilton condition and observing that
\begin{equation}
\frac12\{\abs A^2 -H^2-(R-2\Lam)\}=-\{\ein -\Lam\},
\end{equation}
\cf \cite[equ. (1.1.43)]{cg:cp}, we conclude that the evolution equation
\begin{equation}
\begin{aligned}
(n-1)(H\f^\frac12)'&=-(n-1)\tilde\D w \f^\frac12+(n-1)(R-2\Lam)\f^\frac12 w\\
&\q-2\Lam \f^\frac12 w+\frac{n-1}2H^2\f^\frac12 w
\end{aligned}
\end{equation}
is equivalent to the Hamilton condition provided the tangential Einstein equations are valid.

Finally, expressing the time derivative on the left-hand side by the Poisson brackets such that
\begin{equation}\lae{6.22.1}
\begin{aligned}
(n-1)\{H\f^\frac12,\mc H\}&=-(n-1)\tilde\D w \f^\frac12+(n-1)(R-2\Lam)\f^\frac12 w\\
&\q-2\Lam \f^\frac12 w+\frac{n-1}2H^2\f^\frac12 w
\end{aligned}
\end{equation}
we conclude that the Hamilton equations and the geometric evolution equation \re{6.22.1} are equivalent to the full Einstein equation, \cf the proof of \frt{3.1}.

Switching to the gauge $w=1$ we then quantize the equation \re{6.22.1}. Because of the relation \re{6.16.1} the left-hand side of \re{6.22.1} is transformed to
\begin{equation}
i[\hat H,\f^{-\frac12}\hat g_{ij}\hat\pi^{ij}]=[\hat H,\f^{-\frac12}g_{ij}\frac\de{\de g_{ij}}],
\end{equation}
where $\hat H$ is the transformed Hamiltonian. On the other hand,
\begin{equation}\lae{6.25.1}
\begin{aligned}
\f^{-\frac 12}g_{ij}\frac{\de}{\de g_{ij}}&=\sqrt{n(n-1)}\nu^a D_a=\sqrt{n(n-1)}\nu^0 D_0\\
&=\frac n4\frac{\pa}{\pa t},
\end{aligned}
\end{equation}
where $\nu^a$ is the future unit normal of the hypersurfaces 
\begin{equation}
M(t)=\{\xi^0=t\},
\end{equation}
i.e., the left-hand side of \re{6.25.1} is a constant multiple of the covariant derivative with respect to $t$ in the fiber when the differential operator is applied to functions $u=u(x,g_{ij})$. Hence,
\begin{equation}
\begin{aligned}
&\msp[100]\hp{}[\hat H,\f^{-\frac 12}g_{ij}\frac{\de}{\de g_{ij}}] u\\
&=\f^{-\frac12}g_{ij}\frac\de{\de g_{ij}}\{(R-2\Lam)u\f\}-\f^{-\frac12}(R-2\Lam)\f g_{ij}\pde u{g_{ij}}\\
&= \f^{-\frac12}\{\frac n2(R-2\Lam) u\f -Ru\f -(n-1)\tilde \D u \f\},
\end{aligned}
\end{equation}
in view of \fre{4.47}. The transformation of the right-hand side of \re{6.22.1}, note that $w=1$, yields
\begin{equation}
(n-1) (R-2\Lam)u\f^{\frac12} -2\Lam u\f^\frac12+\f^\frac12\frac{n-1}2H^2 u,
\end{equation}
where
\begin{equation}
\begin{aligned}
\f^\frac12\frac{n-1}2H^2 u&= -\frac n2\f^{-\frac12}\{\frac1{n(n-1)}\f^{-1} g_{ij}g_{kl}\frac \de{\de g{ij}}\frac \de{\de g_{kl}}\}u\\
&= -\frac n2\f^{-\frac12}(\nu^a\nu^bD_aD_b u)
\end{aligned}
\end{equation}
or
\begin{equation}
\f^\frac12\frac{n-1}2H^2 =-\frac n2\f^{-\frac12}D_a(\nu^a\nu^bD_b u)
\end{equation}
depending on the ordering of the  derivatives.

Observing that
\begin{equation}
\nu=(\nu^0,0,\ldots,0)
\end{equation}
and
\begin{equation}
\nu^0=\frac14\sqrt{\frac n{n-1}}
\end{equation}
we obtain, after multiplying both sides with $\f^\frac12$, the hyperbolic equations
\begin{equation}
\frac1{32}\frac {n^2}{n-1}\Ddot u -(n-1)t^2\tilde\D u  -\frac n2 Rt^2u+n\Lam t^2u=0
\end{equation}
or
\begin{equation}
\frac1{32}\frac {n^2}{n-1}t^{-m}\frac{\pa}{\pa t}(t^m\dot u) -(n-1)t^2\tilde\D u  -\frac n2 Rt^2u+n\Lam t^2u=0
\end{equation}
where we recall that $\f=t^2$, \cf \re{4.58} and \fre{4.74}.
 
 These equations can be rewritten, as before, by observing that
 \begin{equation}
g_{ij}=t^{\frac4n}\s_{ij},
\end{equation}
such that
\begin{equation}
\tilde \D u=t^{-\frac4n}\tilde \D_{\s_{ij}}u
\end{equation}
and
\begin{equation}
R=t^{-\frac4n}R_{\s_{ij}},
\end{equation}
where $R_{\s_{ij}}$ is the scalar curvature of the metric $\s_{ij}$. Both equations are hyperbolic equations in $E$, where $u=u(x,t,\xi^A)$, $1\le A\le m$, and $\s_{ij}=\s_{ij}(x,\xi^A)$. However, for fixed $(\xi^A)$, we may consider these equations as hyperbolic equations in
\begin{equation}
\so\times \R[*]_+,
\end{equation}
where the solutions as well as the metric depend on an additional parameter $(\xi^A)$. To simplify the notation let us drop the tilde over the Laplacian and stipulate that the Laplacian as well as the scalar curvature refer to the metric $\s_{ij}$. Then we can rewrite the equations as
\begin{equation}\lae{6.15}
\frac1{32}\frac{n^2}{n-1}\Ddot u
-(n-1) t^{2-\frac4n}\D u-\frac {n}2 t^{2-\frac4n}Ru+nt^2\Lam u=0.
\end{equation}
and
\begin{equation}\lae{6.14}
\begin{aligned}
\frac1{32}\frac{n^2}{n-1}t^{-m}\pde{(t^m\dot u)}t
-(n-1) t^{2 -\frac4n}\D u-\frac {n}2t^{2-\frac4n}Ru+nt^2\Lam u=0
\end{aligned}
\end{equation}
We also note that 
\begin{equation}\lae{6.41.1}
\det \s_{ij}=\det\chi_{ij}
\end{equation}
and that $\s_{ij}\in E_1$ is arbitrary but fixed.
\bl
Both operators are symmetric with respect to the Lorentzian metric
\begin{equation}\lae{6.17}
d\bar s^2=-\frac{32(n-1)}{n^2}dt^2+\s_{ij}dx^idx^j
\end{equation}
and they are normally hyperbolic with respect to the metric
\begin{equation}\lae{6.18}
d\tilde s^2=-\frac{32(n-1)}{n^2}dt^2+\frac1{n-1}t^{\frac4n-2}\s_{ij}dx^idx^j.
\end{equation}
\el
Thus, if 
\begin{equation}
Q=\so\times \R[*]_+
\end{equation}
is globally hyperbolic with respect to these metrics, and if we denote $Q$ equipped with the metric \re{6.18} by $\tilde Q$ and stipulate that $Q$ is equipped with the metric \re{6.17}, then the results from \rs{4} and \rs{5} can be applied to the present setting. 
\bl
Assume that the metric
\begin{equation}
\s_{ij}(x,\xi)\in E_1,
\end{equation}
where $\xi=(\xi^A)$ is fixed, is complete, then the Lorentzian manifolds $Q$ and $\tilde Q$ are globally hyperbolic, and the hypersurfaces
\begin{equation}
M_\tau=\{t=\tau\}\su Q
\end{equation}
are Cauchy hypersurfaces.
\el
\bp
Let us only consider $Q$. From the proof of \frl{4.2} we infer that the claims are correct if a bounded curve
\begin{equation}
\ga(s)\su \so,\qq s\in I,
\end{equation}
where bounded means, bounded relative to $\s_{ij}$, is relatively compact which is the case, if $(\so,\s_{ij})$ is complete.
\ep

In the next theorem we would like to prove that the solutions depend smoothly on $\xi$. In order to achieve this, the Cauchy values have to be prescribed on $E_1(\tau)$ and not only on $M_\tau$.

\bt
Let $P$ be one of the hyperbolic operators in \re{6.14} or \re{6.15}, and let $E_1(\tau)$ be given as well as functions $f\in C^\un_c(E)$ and $u_0,u_1\in C^\un_c(E_1(\tau))$. These functions depend on $(x,t,\xi)$. Since $f,u_0,u_1$ have compact support, the corresponding $\xi$, such that $f(\xi), u_0(\xi), u_1(\xi)$ do not identically vanish in $Q$, are contained in a relatively compact, open set $U$. Assume  that the metrics
\begin{equation}
\s_{ij}(x,\xi),\q\xi\in U,
\end{equation}
are all complete, then, the Cauchy problems
\begin{equation}
\begin{aligned}
Pu&=f\\
\fv u{E_1(\tau)}&=u_0\\
\fv{\dot u}{E_1(\tau)}&=u_1
\end{aligned}
\end{equation}
are uniquely solvable in $(Q,\s_{ij})$ for all $\xi\in U$ such that
\begin{equation}
u=u(x,t,\xi)\in C^\un (\fv EU),
\end{equation}
where
\begin{equation}
\fv EU=\{(x,t,\xi): \xi\in U\}.
\end{equation}
\et
\bp
First, we apply the results in \rs{4} to the operator $P$ and the globally hyperbolic spaces $Q$ and $\tilde Q$ for each $\xi\in U$ to conclude that, for fixed $\xi\in U$, the solutions exist, are uniquely determined, and are smooth in $(x,t)$. Arguing then as in the proof of \cite[Theorem 5.4]{cg:qgravity}, where we considered solutions of hyperbolic problems in the fibers of $E$, where the solutions and the data were depending on the parameter $x\in\so$, we can prove, by considering the problems in $\tilde Q$, so that $P$ is normally hyperbolic, that the solutions are also smooth in $\xi$. Moreover, for each $\xi\in U$, the solution $u(\xi)$ satisfies the known support properties of solutions in $\tilde Q$.
\ep
The equations \re{6.14} or \re{6.15} can be looked at as being gravitational wave equations and the solutions $u=u(x,\xi)$ can be considered to be gravitons. Note that $\xi=(\xi^A)$ are coordinates for the metrics in the fibers, and the pair $(x,\xi)$ represents the metric $\s_{ij}(x,\xi)$ in $\so$.

If $\so$ is compact then we shall construct variational solutions of equation \re{6.15} with finite energy which may be considered to provide a spectral resolution of the problem for fixed $\xi$.

Let us start with the following well-known lemma:
\bl
Let $\so$ be compact equipped with the metric $\s_{ij}=\s_{ij}(\xi)$. Then the eigenvalue problem
\begin{equation}\lae{6.27}
-(n-1)\D v-\frac{n}2R v=\mu v
\end{equation}
has countably many solutions $(v_i,\mu_i)$ such that
\begin{equation}
\mu_0<\mu_1\le \mu_2\le\cdots,
\end{equation}
\begin{equation}
\lim_i\mu_i=\un,
\end{equation}
and
\begin{equation}
\int_\so\bar v_iv_j=\de_{ij},
\end{equation}
where we now consider complex valued functions. The eigenfunctions are a basis for $L^2(\so,\Cc)$ and are smooth.
\el

Now we argue similarly as in \cite[Subsection 6.7]{cg:qfriedman}: Choose any eigenfunction $v=v_i$ with positive eigenvalue $\mu=\mu_i$, then we look at solutions $u$ of \re{6.15} of the form
\begin{equation}
u(x,t)=w(t)v(x).
\end{equation}
Inserting $u$ in the equation we deduce
\begin{equation}
\frac1{32}\frac{n^2}{n-1}\Ddot w +\mu t^{2-\frac4n}w+nt^2\Lam w=0,
\end{equation}
or equivalently,
\begin{equation}\lae{6.33}
-\frac1{32}\frac{n^2}{n-1}\Ddot w -\mu t^{2-\frac4n}w-nt^2\Lam w=0.
\end{equation}
This equation can be considered to be an implicit eigenvalue problem with eigenvalue $\Lam$.

To solve \re{6.33} we first solve 
\begin{equation}\lae{6.34}
-\frac1{32}\frac{n^2}{n-1}\Ddot w +nt^2w=\lam\mu t^{2-\frac4n}w,
\end{equation}
where $\lam$ is the eigenvalue. Let $I=\R[*]_+$ and $H$ be the embedded  subspace of the Sobolev space $H^{1,2}_0(I)$
\begin{equation}
H\hra H^{1,2}_0(I,\Cc)
\end{equation}
defined as the completion of $C^\un_c(I,\Cc)$ under the norm of the scalar product
\begin{equation}
\spd w{\tilde w}_1=\int_I\{\bar w'\tilde w' +t^2\bar w\tilde w\},
\end{equation}
where a prime or a dot denotes differentiation with respect to $t$. Moreover, let $B$, $K$ be the symmetric forms
\begin{equation}
B(w,\tilde w)=\int_I\{\frac1{32} \frac{n^2}{n-1}\bar w'\tilde w'+nt^2\bar w\tilde w\}
\end{equation}
and
\begin{equation}
K(w,\tilde w)=\int_I\mu t^{2-\frac4n}\bar w\tilde w,
\end{equation}
then the eigenvalue equation \re{6.34} is equivalent to 
\begin{equation}\lae{6.39}
B(w,\f)=\lam K(w,\f)\qq\A\,\f\in H
\end{equation}
as one easily checks.
\bl
The quadratic form $K(w)=K(w,w)$ is compact relative to the quadratic form $B$, i.e., if $w_k\in H$ converges weakly to $w\in H$
\begin{equation}
w_k\rha w\qq \text{in } H,
\end{equation}
then
\begin{equation}
K(w_k)\ra K(w).
\end{equation}
\el
\bp
The proof is essentially the same as the proof of \cite[Lemma 6.8]{cg:qfriedman} and will be omitted.
\ep

Hence the eigenvalue problem \re{6.39} has countably many solutions $(\tilde w_i,\lam_i)$ such that
\begin{equation}\lae{6.42}
0<\lam_0<\lam_1<\cdots,
\end{equation}
\begin{equation}
\lim\lam_i=\un
\end{equation}
and
\begin{equation}
K(\tilde w_i,\tilde w_j)=\de_{ij}.
\end{equation}
For a proof of this well-known result, except the strict inequalities in \re{6.42},  see e.g. \cite[Theorem 1.6.3, p. 37]{cg:pdeII}. Each eigenvalue has multiplicity one since we have a linear ODE of order two and all solutions satisfy the boundary condition 
\begin{equation}\lae{6.45}
\tilde w_i(0)=0.
\end{equation}
The kernel is two-dimensional and the condition \re{6.45} defines a one-dimen\-sional subspace. Note, that we considered only real valued solutions to apply this argument. 

Finally, the functions
\begin{equation}
w_i(t)=\tilde w_i(\lam_i^{-\frac n{4(n-1)}}t)
\end{equation}
then satisfy \re{6.33} with eigenvalue
\begin{equation}
\Lam_i=-\lam_i^{-\frac n{n-1}}
\end{equation}
and 
\begin{equation}
u_i=w_iv
\end{equation}
is a solution of the wave equation \re{6.15} with finite energy
\begin{equation}\lae{6.49} 
\begin{aligned}
\norm{u_i}^2=\int_Q\{\abs {\dot u}^2+(1+t^2)\s^{ij}\bar u_iu_j+\mu t^{2-\frac4n}\abs u^2\}<\un.
\end{aligned}
\end{equation}
Note that the actual energy is defined by a weaker norm
\begin{equation}
\int_Q\{\abs {\dot u}^2+t^{2-\frac4n}\s^{ij}\bar u_iu_j+\mu t^{2-\frac4n}\abs u^2\}
\end{equation}
which is of course bounded too.

Let us summarize these results:
\bt\lat{6.7}
Assume $n\ge 2$ and $\so$ to be compact and let $(v,\mu)$ be a solution of the eigenvalue problem \re{6.27} with $\mu>0$, then there exist countably many solutions $(w_i,\Lam_i)$ of the implicit eigenvalue problem \re{6.33} such that
\begin{equation}
\Lam_i<\Lam_{i+1}<\cdots <0,
\end{equation}
\begin{equation}
\lim_i\Lam_i=0,
\end{equation}
and such that the functions
\begin{equation}
u_i=w_i v
\end{equation}
are solutions of the wave equations \re{6.15}. The transformed eigenfunctions
\begin{equation}
\tilde w_i(t)=w_i(\lam_i^{\frac n{4(n-1)}}t), 
\end{equation}
where
\begin{equation}
\lam_i=(-\Lam_i)^{-\frac{n-1}n}
\end{equation}
form a basis of the Hilbert space $H$ and also of $L^2(\R[*]_+,\Cc)$.
\et

\br
Let $\s_{ij}$ be a smooth and complete Riemannian metric in $\so$, then $\s_{ij}$ is in general only a section of $E$ but not an element. However, the metric $\chi_{ij}$ in \fre{4.1}, which we used to define $\f$ in order to  replace the density $\sqrt g$, can certainly be assumed to belong to $E$, and hence to the subbundle $E_1$, because we can easily define a covering of local trivializations where $\chi$ is always part of the generating local frames. Since $\chi$ is chosen arbitrarily we may just as well assume that
\begin{equation}
\chi_{ij}=\s_{ij}.
\end{equation}
Hence, the hyperbolic equations \re{6.15} or \re{6.14}, which are supposed to describe a model for quantum gravity, can be applied to any given smooth and complete metric metric $\s_{ij}$, or more precisely, to any complete Riemannian manifold $(\so,\s_{ij})$.
\er
Let us formulate this result  in case of equation \re{6.15} as a theorem:
\bt
Let $(\so,\s_{ij})$ be a connected, smooth and complete  $n$\ndash di\-mensional Riemannian mani\-fold   and let 
\begin{equation}
Q=\so\times \R[*]_+
\end{equation}
be the corresponding globally hyperbolic spacetime equipped with the Lorentzian metric \re{6.17} or, if necessary, with \re{6.18}, then the hyperbolic equation
\begin{equation}
\frac1{32}\frac{n^2}{n-1}\Ddot u
-(n-1) t^{2-\frac4n}\D u-\frac {n}2 t^{2-\frac4n}Ru+nt^2\Lam u=0,
\end{equation}
where the Laplacian and the scalar curvature correspond to the metric $\s_{ij}$, describes a model of quantum gravity. If $\so$ is compact a spectral resolution of this equation has been proved in \rt{6.7}.
\et
\br
If $\so$ is not compact, then we proved in \cite{cg:uf4,cg:qbh,cg:qbh2} that a spectral resolution is possible if either $\so$ is an asymptotically Euclidean Cauchy hypersurface of a globally hyperbolic spacetime $N$, or, if $N$ is a black hole, if $\so$ is the smooth limit of Cauchy hypersurfaces representing the event horizon though with a different metric.
\er
\br
When $\s_{ij}$ is the metric of a space of constant curvature then the equation \re{6.15}, considered only for functions $u$ which do not depend on $x$,  is identical to the equation obtained by quantizing the Hamilton constraint in a Friedmann universe without matter but including a cosmological constant. The equation is the ODE
\begin{equation}
\frac1{16} \frac n{n-1} \Ddot u-R r^{2-\frac 4n}u+2r^2\Lam u=0,\qq 0<r<\un,
\end{equation}
\cf \cite[equ. (3.37)]{cg:qfriedman}, though the equation there looks differently, since in that paper we divided the Lagrangian by $n(n-1)$.
\er

\bibliographystyle{hamsplain}
\providecommand{\bysame}{\leavevmode\hbox to3em{\hrulefill}\thinspace}
\providecommand{\href}[2]{#2}



\end{document}